\documentclass[twocolumn]{svjour3}         
\smartqed  
\usepackage{graphicx}
\begin{document}

\title{Observational diagnostics of gas in protoplanetary disks
}


\author{Andr\'es Carmona         
}


\institute{Andr\'es Carmona \at
              ISDC Data Centre for Astrophysics  \& Geneva Observatory,\\ 
              University of Geneva \\
              chemin d'Ecogia 16 \\
              1290 Versoix, Switzerland\\ 
              \email{andres.carmona@unige.ch}           
}

\date{October 2008.\\ 
 Review written for the proceedings of the conference "Origin and Evolution of Planets 2008", Ascona, Switzerland, 2008. }

\maketitle

\begin{abstract}
Protoplanetary disks are composed primarily of gas (99\% of the mass).
Nevertheless, 
relatively few observational constraints exist
for the gas in disks.
In this review,
I discuss several observational diagnostics 
in the UV, optical, near-IR, mid-IR, and (sub)-mm wavelengths
that have been employed to study the 
gas in the disks of young stellar objects.
I concentrate in diagnostics that probe the
inner 20 AU of the disk, the region where planets are expected to form.
I discuss the potential and limitations of each gas tracer
and present prospects for future research.

\end{abstract}

\section{Introduction}
\label{intro}
At the time when giant planets form, 
the mass (99\%) of the protoplanetary disk is dominated by gas.
Dust is a minor constituent of the mass of the disk.
Still, it dominates the opacity; consequently, it is much easier to observe.
Therefore, most of the observational constraints of disks
have been deduced from the study of dust emission (e.g., see chapter by Henning et al.).
In contrast, direct observational constraints of the gas in the disk
are relatively scarce.
Nonetheless, to obtain direct information from the gas content
of the disk is crucial for answering fundamental questions in planet
formation such as:
how long does the protoplanetary disk last?,
how much material is available for forming giant planets?,
how do the density and the temperature of the disk vary as a function of the radius?,
what are the dynamics of the disk?.

Although we have learned important insights from disks from dust observations (see for example
the reviews by Henning et al. 2006 and Natta et al. 2007),
dust presents several limitations: 
(i) dust spectral features are broad; consequently, dust emission does not
provide kinematical information;
(ii) dust properties are expected to change during the planet formation 
process;
therefore, quantities such as the gas-to-dust ratio (needed to derive 
the disk mass from dust continuum emission in the (sub)-mm) 
are expected to strongly vary with respect to the
conditions of the Interstellar Medium (ISM).
In addition, dust signatures are strongly related to dust size. 
In particular, as soon a dust particle reaches a size close to a decimeter 
it becomes practically "invisible".
For example, a crucial quantity such as 
the disk dissipation time scale is normally deduced from the decline
of the fraction of the sources presenting near-IR (JHKL) 
excess in clusters of increasing ages (e.g., Haisch et al. 2001). 
In reality, the time scale that we want to constrain 
is not the time scale in which the small warm dust particles disappear from the surface layers in the inner 
disk, but the time scale in which the gas in the disk disappears.

In summary, independent information of the gas is crucial to understand protoplanetary 
disk structure and {\it combined studies of gas and dust in the disk are needed}. 
In particular, we are interested in deriving constraints for the inner disk,
the region where the planets are expected to form (R$<$20 AU).
The observational study of the gas in the inner disk is a relatively new 
emerging topic.
Only with the advent of ground-based high-spectral resolution infrared
spectrographs and space-born infrared spectrographs has this research become possible.

\begin{figure}
\includegraphics[width=0.5\textwidth]{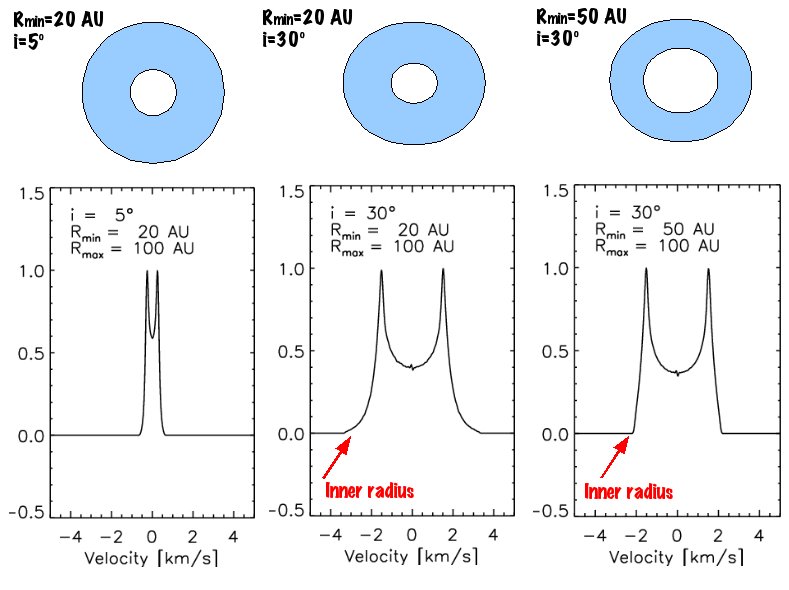}
\caption{Line shape, inclination and the emitting region of a disk
in Keplerian rotation. 
The double peaked profile indicates emission from a rotating disk.
The line width and double peak separation depend on the inclination 
of the disk. 
The line profile wings depend on the innermost radius of the emission.
In this toy model, it is assumed that the intensity is constant as a function of 
the radius and a spectral resolution of 1 km/s.}
\label{fig:1}       
\end{figure} 

The main tool used to study the gas in the disk is molecular spectroscopy.
The gas in the disk is heated by collisions with dust, shocks or UV or  X-rays
from the central star. The heated molecules populate different rotational and vibrational 
levels and when they de-excite emission lines are produced.
The key is that these emission lines have the imprint of 
the physical conditions of the gas where they originated. 
Important constraints can be derived from emission lines:
(i) line ratios of different transitions constrain the excitation mechanism
(shocks, UV or X-rays) and the temperature of the gas responsible of the emission;
(ii) if the emitting gas is optically thin, then the measured line fluxes are proportionally 
related to the amount of molecules present; therefore, line fluxes can be used to 
derive column densities and gas masses;
(iii) line shapes and spatial extended provide constraints on the size and
geometry of emitting region as well as the dynamics of the emitting gas.
Figure 1 displays a pedagogical example of how a line profile
changes as a function of the inclination and size of the emitting region
in the disk. 
The line width and double peak separation depend on the inclination . 
The line profile wings depend on the innermost radius of the emission.
For the interested reader, 
recent reviews  by Najita et al. (2000 \& 2007a), Blake (2003) and Carr (2005)
cover the subject of observational studies of gas in disks.

\begin{figure}
\includegraphics[width=0.5\textwidth]{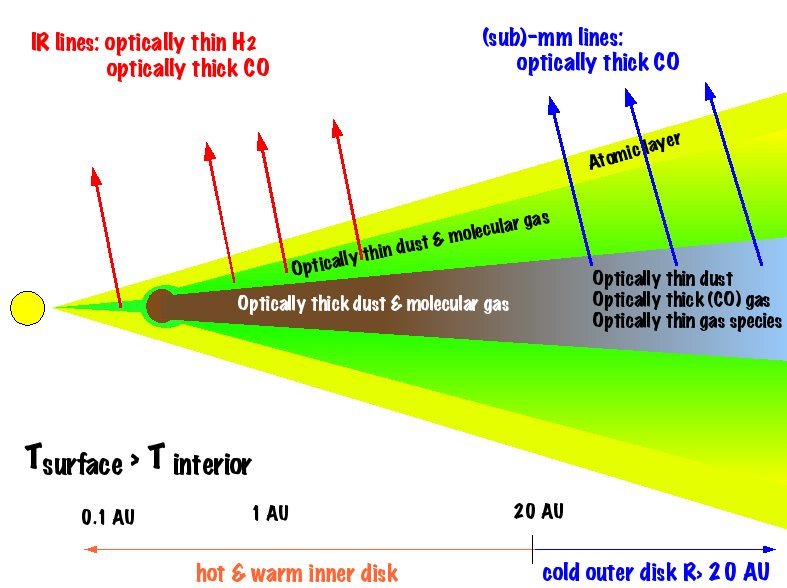}
\label{fig:2}
\caption{Cartoon of the structure of an optically thick disk.
The disk has a hot inner region (R$<$20 AU) and a cold outer
region (R$>$20 AU). Near-IR and mid-IR diagnostics probe the inner disk, 
(sub)-mm diagnostics probe the outer disk. 
The disk has a vertical structure: a dense mid-plane and a hotter 
less dense surface layer.
Near-IR and mid-IR gas and dust emission features originate in the optically thin 
surface layer of the inner disk. Near-IR and mid-IR continuum originates from the optically
thick interior layer. 
At (sub)-mm wavelengths we observe line emission from optically thick CO gas  
and optically thin emission of cold dust located in the outer disk.}
\end{figure}

\section{The gas emitting region.}
An important element that one should always bear in mind
when studying molecular emission from disks
is the region where the emission observed is produced.
As the disk has a radial and vertical temperature and density gradient 
(hotter closer to the star and higher in the disk, denser closer to the star and deeper in the disk)
we need to have clear three concepts:
(i) inner and outer disk,
(ii) interior and surface layer,
(iii) optical depth of the medium (i.e., optically thin or thick) and its wavelength 
and molecular species dependency. 

The first concept is related to the distance of the central star (i.e., the radius)
and the radial temperature gradient.
Different gas tracers probe different radial regions of the disk.
The hot and warm (T$\sim$100-2500 K) {\it inner disk} (R$<$20 AU) is traced by lines
in the UV, near- and mid-IR.
The cold  (T$<$100 K) {\it outer disk} (R$>$20 AU) is probed by lines in the (sub)-mm. 

The second and third concepts are related to the vertical temperature and density gradient.
As a first approximation, we can understand a protoplanetary disk at each radius 
as composed of two vertical layers: 
a dense colder {\it interior layer} and less dense hotter {\it surface layer} (Chiang and Goldreich 1997).
At near-IR and mid-IR wavelengths (that probe R$<$20 AU),  
the dust and gas in the interior layer are optically thick and the interior layer
radiates as a black body (i.e., continuum equal to the blackbody level).
Gas lines -and dust emission features- in the UV, near-IR and mid-IR arise from the optically thin
surface layer of the inner disk. They and are observed on top of the continuum of the interior layer. 
Emission is observed from the surface layer because the surface layer is hotter
than the interior layer. 
The gas emission from the surface layer could be optically thin or optically thick
depending on the gas density and the molecular species (for example CO emission is generally optically thick,
and H$_2$ emission is optically thin).
Optically thick gas lines from the surface layer of the inner disk are observable because the dust in the surface 
layer is optically thin. 

At (sub)-mm wavelengths (that probe R$>$20 AU),
the dust become optically thin in the interior layer as well (i.e., continuum below the black-body level),
and we can observe optically thick emission from - CO - gas from the interior layer on 
top of the optically thin dust continuum. Optically thin gas emission can be observed from other less abundant
molecular species such as $^{13}$CO. 
Figure 2 presents a cartoon
of the structure of a disk and the observable properties of the medium.  

In addition to the surface layer, 
in particular cases such as transitional disks\footnote{Transitional disks are
sources that do not display or display very weak near-IR excess but that 
have relatively normal mid-IR excess in their spectral energy distributions (SED).
This is due to the lack of small hot dust particles radiating in the 
near-IR.
This can be attributed to a hole in the disk (a physical gap, i.e., no gas and no dust), 
to a lack of dust particles in the inner disk (i.e., no dust, but gas is still present), 
or a change in the dust size
(an opacity gap, i.e., gas and dust are still there but the dust is too large to radiate in the near-IR).
Gaps could be produced either by a close binary, 
a low-mass companion (i.e., a giant planet),
dust coagulation, 
or photoevaporation of the inner disk. For recent papers on transitional disks,
see for example Calvet (2005), Sicilia-Aguilar et al. (2006), Najita et al. (2007b), Pontoppidan et al. (2008) and Cieza et al. (2008).}
or disks in close binaries,
gas emission lines (optically thin or thick depending on the gas density)
are expected from the inner region of the disk when the dust is optically thin. 

\begin{figure}
\includegraphics[width=0.5\textwidth]{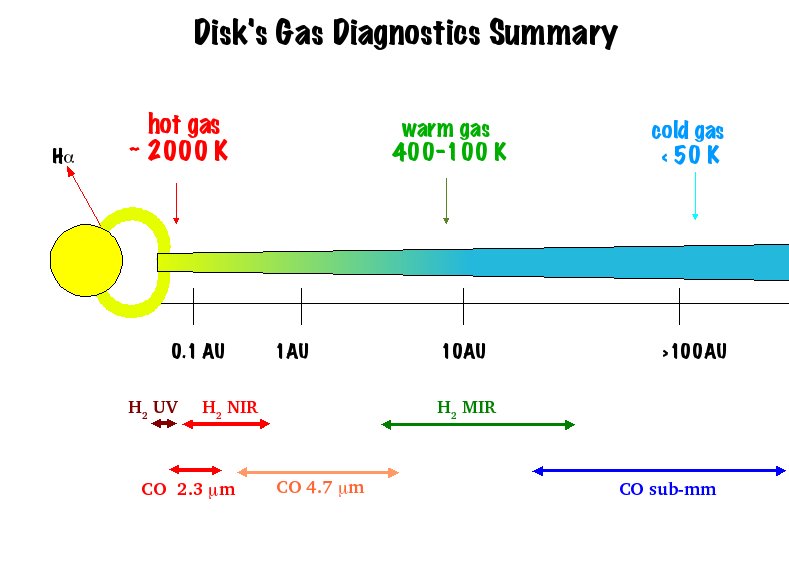}
\caption{Summary of disk gas diagnostics and 
the disk region that they probe.}
\label{fig:3}
\end{figure}

\section{A diversity of gas diagnostics}
Given that the disk has a radial temperature gradient, 
different tracers are employed to probe the gas at each location 
in the disk.
Figure 3 displays a summary of disks gas diagnostics.
In the following sections, 
I will explain one by one each of them highlighting their strengths and limitations.
I will follow a discussion inspired by the historical development
of the field.

\subsection{H$\alpha$ in emission}
Properly speaking H$\alpha$ in emission (6590 \AA) from young
stars  does not trace the gas in the disk.
However, since H$\alpha$ emission is one of the primary indicators of
accretion of gas from the disk onto the star (e.g., Hartmann et al. 1994;
Muzerolle et al. 1998b),
H$\alpha$ is a robust indicator of a gaseous disk.
In particular, when any other gas tracer fails, 
H$\alpha$ emission tells us unambiguously that there is gas.

Its main advantages are:
(i) it is easy to observe
(a small telescope with a medium resolution R$\sim$3000 spectrograph can do it);
(ii) it is a robust diagnostic; 
(iii) it can be studied in large samples
of objects of diverse brightness in relatively short time.

Its main limitations are:  
(i) it does not provide information 
on the physical conditions of the gas in the disk 
(H$\alpha$ tells us that there is gas and allows us to deduce the accretion rate, but not much more);
(ii) if the accretion rate is below 10$^{-12} $M$_{\odot}$/yr,
H$\alpha$ in emission is too weak to be observed;
(iii) in low resolution spectra of late type stars (K and later), 
H$\alpha$ in emission from accretion can be confused with H$\alpha$ due to stellar activity
(typically a line width higher than 100 km/s is required for inferring accretion).
H$\alpha$ in emission allowed us to confirm that very low mass stars have disks,
that relatively old objects still have gaseous disks,
and that several transition disks 
still have gas in their inner disks (e.g., Sicilia-Aguilar et al. 2006).

Note that other H lines such as the weaker Br$\gamma$ line in the near-IR
have been used for deducing accretion rates -and the presence of a gaseous disk- 
in young stars (e.g., Najita et al. 1996a, Muzerolle et al. 1998a,
Natta et al. 2006).

\subsection{CO overtone emission emission at 2.3 $\mu$m}
One of the earliest indicators of gas in a disk was the detection of 
CO overtone ($\Delta \upsilon$ = 2) emission at 2.3 $\mu$m in T Tauri stars 
(Scoville et al. 1983; Geballe and Persson 1987; Carr et al. 1989, 1992 \& 1993; Chandler 1993; Najita et al. 1996b).
More recently Thi et al. (2005a) %
reported CO 2.3 $\mu$m emission in the Herbig Ae/Be star 51 Oph
(see also Berthoud et al. 2007 and Tatulli et al. 2008).
CO bandhead emission traces very hot (T$>$2000 K) and dense gas in the
innermost part of the disk (R$<$0.1 AU) in objects exhibiting high accretion.
This line provided the first evidence of Keplerian rotation of gas in
the inner disk of T Tauri stars. 
The CO band-head emission is characterized by a broad feature at 2.3 $\mu$m.
Its particular shape can be modeled by the convolution of  
$\upsilon$=2-0 CO ro-vibrational emission lines at rest with the double peaked 
line profile characteristic of emission from a disk.

The main advantages of this diagnostic are: (i) it can be studied with a low resolution
near-IR spectrographs and (ii) it traces the innermost part of the disk.
The main disadvantage is that the densities and temperatures require to be very high
to produce the emission. This is the main reason for why this line
has been observed towards only a hand-full of objects.
Finally, note that in general the CO overtone emission at 2.3 $\mu$m is
optically thick emission. 
For the interested reader this diagnostic and its modeling
is treated in detail in Carr et al. (1993).

\subsection{CO (sub)-mm emission and other (sub)-mm lines}
Progress in the sensitivity of millimeter radio telescopes led to the detection
in the early 90's  of pure rotational emission ($\Delta J=1$) from cold CO from T Tauri stars 
(e.g., Weintraub et al. 1989; Koerner et al. 1993; Skrutskie et al. 1993;
Beckwith \& Sargent 1993; Guilloteau \& Dutrey 1994).
Those observations revealed the classical double peaked profile 
expected from gas rotating in Keplerian disk (see Figure 1).
The detections with single dish telescopes were followed by studies with
millimeter interferometers that allowed to spatially resolve the emission in T Tauri stars
(e.g., Dutrey et al. 1996; Guilloteau \& Dutrey 1998; Dutrey et al. 1998)
and Herbig Ae/Be stars (e.g., Mannings \& Sargent 1997).
In the particular case of Herbig Ae stars,
millimeter interferometry observations of CO permitted to
settle the controversy about the existence of circumstellar disks
around these objects.
Since then a multitude of studies in the (sub)-mm have been undertaken
in protoplanetary disks thanks to development of sensitive millimiter and sub-mm arrays 
(e.g., Plateau de Bure, SMA, CARMA, see recent review by Dutrey et al. 2007).
Progress has not only been done in tracing cold CO, but
also in detecting many other molecules such as CN, HCN, HCO$^+$, H$_2$CO, etc (e.g., Thi et al. 2004).
These detections have opened the door to observational studies of chemistry in disks 
(e.g., review by Bergin et al. 2007).

The principal advantages of CO are:
(i) CO is the second most abundant molecule after the H$_2$. 
(ii) CO (sub)-mm emission is bright. 
These not trivial characteristics had allowed the survey
of relatively large samples of young stars with disks (e.g., Dent et al. 2005, Andrews \& Williams 2005),
and have permitted the spatially resolved study of the disk structure in several objects (e.g., Pi\'etu et al. 2007).         
The main limitation of CO emission and other gas tracers in the (sub)-mm
is that they trace the cold gas in the outer part of the disk (R$>$20 AU).
Therefore, they are not suitable to trace the disk region (R$<$20 AU)
where planets are expected to form.
Nonetheless, most of the mass of the disk is located in the cold outer part of
the disk.

CO emission in the (sub)-mm has important limitations as tracer of gas mass: 
(i) cold CO is expected to freeze-out onto dust grain surfaces at low temperatures (T$<$20 K) and
high densities (N$\geq10^{5}$ cm$^{-3}$); therefore, CO is expected to be depleted;
(ii) CO emission can be optically thick.
Both limitations introduce a large uncertainty on the conversion factor (up to a 1000) 
between the CO observed and the real amount of H$_2$ present in the outer disk.
Typically, the ISM CO/H$_2$ conversion factor of 10$^{-4}$ is used.
However, this value is most likely incorrect for disks. 
The large uncertainty in the determination of mass of gas affects seriously 
the empirical estimation of the gas-to-dust ratio.
Finally, note that there are sources which are known to have a disk
 - by the detection of IR \& (sub)-mm excess and H$\alpha$ in emission -, 
but, because of sensitivity limitations or depletion,  
we do not detect CO emission.

In the near future, 
significant progress is expected in the study of disks in the (sub)-mm domain
thanks to the Atacama Large Millimiter/submillimiter Array (ALMA), which will start operations soon.
ALMA will increase the sensitivity and spatial resolution 
by an order of magnitude with respect to those obtained with present facilities.
In addition, the far-infrared space telescope HERSCHEL will
allow the study of disks at frequencies which are impossible 
to observe from the ground. 
A detailed description of gas diagnostics in the (sub)-mm and chemistry in disks
deserves a review in its own. For the interested reader, 
the reviews by Dutrey et al. (2007) and Bergin et al. (2007) cover these topics.

\begin{figure}
\includegraphics[width=0.5\textwidth]{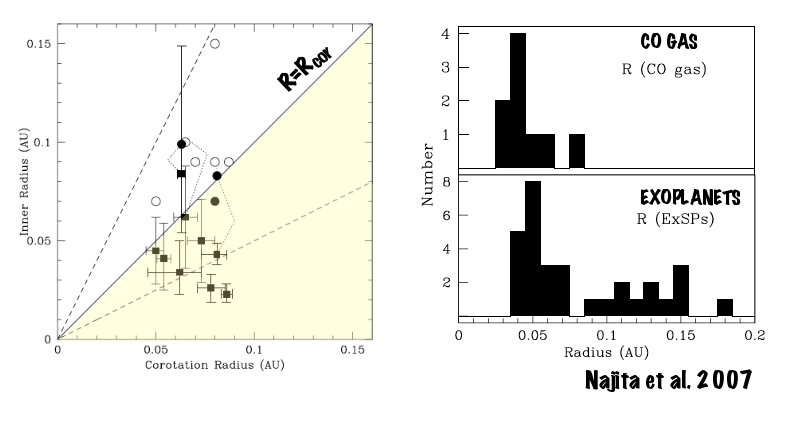}
\caption{{\it Left panel:} Inner radius of CO fundamental emission at 4.7 $\mu$m
vs corotation radius of the sources. In black squares CO 4.7 $\mu$m detections (Najita et al. 2007a),
in filled circles the inner dust radius from near-IR interferometry (Akeson et al. 2005a, b.),
in open circles R$_{\rm in}$ from SED modeling (Muzerolle et al. 2003).
In several sources CO gas is present inside the corotation radius and inward the dust sublimation radius. 
{\it Right panel:} Distributions of the gas inner radius from CO 4.7 $\mu$m observations and
the distribution of the semi-mayor axis of short period extrasolar planets. 
The minimum radius of both distributions is similar. Both figures are taken from the review from 
Najita et al. (2007a).}
\label{fig:4}
\end{figure}

\subsection{CO emission band at 4.7 $\mu$m}     
The advent of high-resolution (R$>$10000) spectrographs in the near-IR
opened the door to the study of the CO ro-vibrational emission 
band ($\Delta \upsilon=1$) at 4.7 $\mu$m.
CO emission at 4.7 $\mu$m probes the gas at temperatures ranging 
from hundred to thousand of degrees.
This diagnostic is important because the
gas at these temperatures is located  
in the terrestrial planet forming region of the
disk (R $<$5 AU).
CO 4.7 $\mu$m emission from disks has been studied 
in T Tauri stars (e.g., Carr et al. 2001; Najita et al. 2003;  Brittain et al. 2005 \& 2007b; 
Rettig et al. 2005; Brown et al. 2005), 
Herbig Ae/Be stars (e.g., Brittain et al. 2002, 2003 \& 2007a; Blake \& Boogert 2004; 
Carmona et al. 2005; Goto et al. 2006; van der Plas et al. 2008b) 
and in a growing number of 
transitional disks (Rettig et al. 2004; Salyk et al. 2007; Najita et al. 2008; Pontoppidan et al. 2008).

The detection of the CO band at 4.7 $\mu$m allowed in the
first place to constrain the temperature of the gas by means of rotational diagrams.
The gas temperatures observed varied from 100 K to 1000 K and in a few cases up to 3000 K.
In several sources, for example AB Aur (Brittain et al. 2003),
the shape of the rotational diagram allowed to infer CO gas at two different temperatures.
The shape of the CO lines was used to set constraints on the inclination of inner disks (e.g., Blake \& Boogert 2004).
In general, within the errors, the inclinations deduced from CO 4.7 $\mu$m emission 
were consistent with the inclinations derived from (sub)-mm observations (note that to derive the disk
inclination, the CO emitting region needs to be assumed).

Modeling of the shape of the line profile allowed to estimate the innermost radius
where the CO emission is produced (see Figure 1).
Najita et al. (2007a) found that the inner radius of the CO 4.7 $\mu$m emission in 
T Tauri stars extends inside the dust sublimation radius and, typically, inside the corotation radius too.
Those authors compared the distribution of the inner radii deduced from the CO 4.7 $\mu$m lines
and the distribution of the orbital radii of short period extrasolar planets.
They found that both distributions are very similar (see Figure 4).
The gas required for giant planet migration was put in evidence by the CO fundamental transitions.  

\begin{figure}
\includegraphics[width=0.5\textwidth]{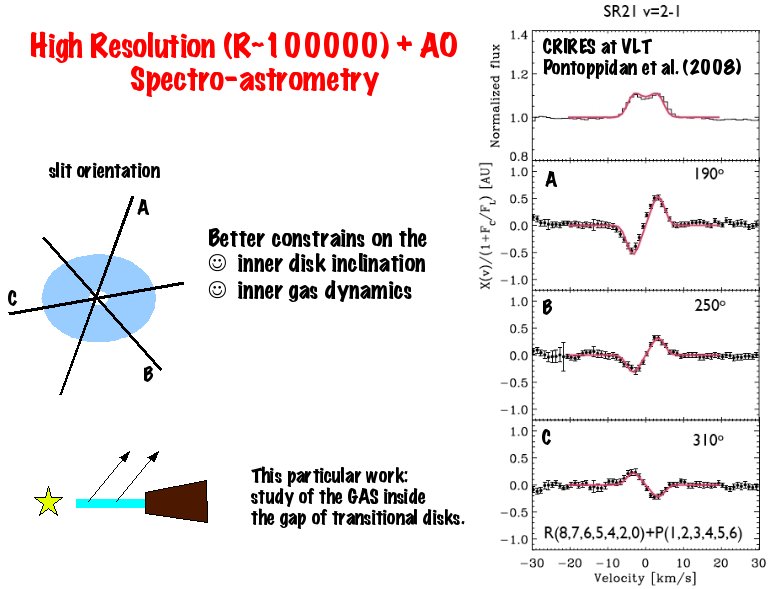}
\caption{Normalized flux and spectro-astrometric signal of the
averaged CO $\upsilon$=2-1 emission at 4.7 $\mu$m from the
transition disk SR 21. 
Data in three slit orientations is displayed.
In the bottom of the right panel, 
a detail of the transitions (R and P) employed for
calculating the composite profile is shown. 
In all the panels the continuous line is a Keplerian disk model.
The line shape and spectro-astrometric signal observed are consistent
with CO gas in Keplerian rotation. Adapted from Pontoppidan et al. (2008). }
\label{fig:5}       
\end{figure}

The most recent development has been to combine  
high spectral resolution with
the enhanced spatial resolution provided by Adaptive Optics (AO).
Goto et al. (2006) spatially resolved the CO $\upsilon$=2-1 4.7 $\mu$m emission from the Herbig Ae/Be
star HD 141569 and measured an inner clearing of radius 11$\pm$2 AU .
Pontoppidan et al. (2008) and van der Plas et al. (2008)
observed young stars with disks combining AO and very high spectral resolution 
(R$\sim$100000). Those authors, employing the spectroastrometry technique,
were able to measure the displacement of the PSF center as a function of wavelength in the line,
and obtained robust evidence of emission from Keplerian disks (see Figure 5).
Pontoppidan et al. (2008) detected gas emission from the inner gaps of transitional disks
(Sr 21, HD 135344B an TW Hya).
  
Besides the fact that CO 4.7 $\mu$m emission probes the terrestrial planet
forming region of the disk, one clear advantage of CO 4.7 $\mu$m  emission 
is that it is detectable in a large number of sources with present instrumentation.
This opens the perspective for developping large surveys.
For example, CO 4.7 $\mu$m emission has been detected in a vast majority of Herbig Ae/Be stars that
display excess $E(K-L)>$1 (Brittain et al. 2007). 
CO 4.7 $\mu$m emission studies have in consequence an important potential for the future.
A second advantage comes from the fact that several lines are observable at the same time 
in a single setup even with echelle spectrographs. 
This is important to make reliable rotational diagrams
and to perform analyses involving the averaging of line profiles (i.e., determine R$_{\rm in}$ or the inclination).

The principal limitations of this diagnostic are: 
(i) it requires high-spectral resolution. Thus it is limited to relatively bright targets 
(for the moment stars fainter than M=10 are challenging) ;
(ii) observations in the M band are relatively time intensive and observations of telluric standard stars are required
before or after the science observations;
(iii) CO 4.7 $\mu$m lines lie very close, if not 
inside CO atmospheric features. Hence, observations should be performed in an epoch such that 
the velocity shift due to the orbital motion of the Earth is maximized;
(iv) one should keep in mind that CO 4.7 $\mu$m emission does not trace in general (unless we see emission from a 
gap) the giant planet forming region of the disk (R$>$5-10 AU).

\subsection{H$_2$ emission from disks}

From previous sections, the reader perhaps 
realizes that principally CO has been discussed .
The discussion started with CO at 2.3 $\mu$m, 
it was followed with CO in the (sub)-mm,
and finally CO emission at 4.7 $\mu$m was treated.
But what about molecular hydrogen?
H$_2$ is by far more abundant 
than CO (in the ISM H$_2$/CO is typically 10$^4$).
Although H$_2$ is the principal constituent of the gas in the disk,
it is very challenging to detect.
This is due to the H$_2$ physical nature:
H$_2$ is an homonuclear molecule that
- in contrast to CO - has not permanent dipole moment. 
Its fundamental transitions are quadrupole in nature. 
Hence, their Einstein spontaneous emission
coefficients are small and give rise to weak lines.
In the context of circumstellar disks, 
one extra challenge is present.
The weak H$_2$ lines should be detected on top of the strong dust continuum emission.
Thus, high spectral resolution is a must to disentangle the weak lines.
To complicate a little bit the life of astronomers,
the energy levels of H$_2$ are widely spaced in energy 
(for example the first  energy level starts at 510 K).
That means that a dedicated observation is required for each H$_2$ line
in order to construct rotational diagrams
(note that in contrast, CO lines are closely spaced and several lines are observable within one
observation set up).
Moreover, since H$_2$ emission is produced in warm T$>150$ K gas,
H$_2$ emission can also originate from shocked gas in outflows and not in the disk
(note that historically H$_2$ emission has been used to trace shocks).
Finally, the mid-IR region is a challenging window to perform high-sensitivity
observations from the ground.

Nevertheless, besides being the most abundant molecule, H$_2$ has some advantages: 
(i) H$_2$ is optically thin up to column densities of 10$^{23}$ cm$^{-2}$; 
thus, if the observed emission is thermal,
we can derive direct constrains to the mass of the emitting gas;
(ii) H$_2$ self-shields against photo-dissociation;
(iii) H$_2$ traces the inner disk in the giant planet formation region;
(iv)  the H$_2$ condensation temperature is 2 K. Hence, H$_2$
does not freeze onto dust grains surfaces -as happens with CO-;
(v) the velocity shift of H$_2$ lines and their spatial extent allow
to distinguish if the emission observed is from a disk or from an outflow.

To discuss the observations of H$_2$ emission in disks, 
I will start by addressing the near-IR ro-vibrational lines,
later I will discuss the mid-IR pure rotational lines and 
finally I will discuss the H$_2$ electronic transition lines in the UV.

\subsubsection{H$_2$ ro-vibrational emission at 2 $\mu$m}
H$_2$ lines in the near-IR probe the inner disk from a fraction of AU up to a few AU.
They trace gas at temperatures of 1000 K and higher.
They are sensitive up to a few lunar masses of gas.
H$_2$ near-IR emission arises from ro-vibrational transitions
between the rotational levels (J) of the first and second vibrational states 
($\upsilon$=1,2) and the rotational levels of the ground vibrational state ($\upsilon$=0).
Typical lines observed are the $\upsilon$=1-0 S(1) line (J=3-1) at 2.12 $\mu$m, 
the 1-0 S(0) line at 2.22 $\mu$m, and the 2-1 S(1) line at 2.24 $\mu$m.

The H$_2$ 1-0 S(1) line was first observed in outflows in T Tau stars 
(e.g., Beckwith et al. 1978; van Langevelde et al. 1994).
Only recently, H$_2$ 1-0 S(1) quiescent emission (i.e., at the velocity of the star) from a disk 
was reported in the T Tauri star TW Hya (Weintraub et al. 2000).
Later on Bary et al. (2002) detected the same line in the weak line T Tauri 
star\footnote{Weak-line T Tauri stars are stars that do not exhibit signatures of
active accretion and that lack IR-excesses in their spectral energy distribution. 
They were discovered by X-ray studies of star-forming regions.}(WTTS) DoAr 21. 
After these initial studies, quiescent H$_2$ 1-0 S(1) emission from T Tauri stars
has been reported by Bary et al. (2003), Itoh et al. (2003), Weintraub et al. 2005,
Ramsay Howat \& Greaves (2007), Carmona et al. (2007, 2008a) and Bary et al. (2008).
Carmona et al. (2007) detected for the first time the H$_2$ 1-0 S(0) 
line in the disk of LkH$\alpha$ 264 and demonstrated empirically that H$_2$ near-IR from disks 
originates in gas at temperatures around 1000 K.

\begin{figure}
\includegraphics[width=0.45\textwidth]{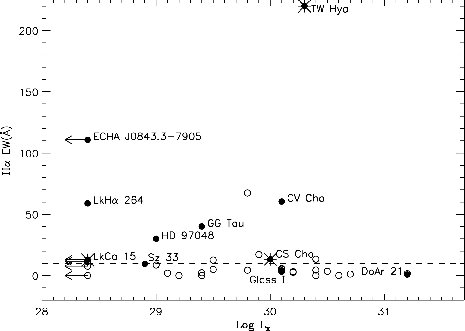}
\caption{H$\alpha$ Equivalent Width vs X-ray luminosity in
CTTS and WTTS in which 1-0 H$_2$ S(1) emission at 2.12 micron has been
detected. Filled circles represent detections, empty circles non-detections,
left arrow upper limits in X-ray luminosity, asterisks transition disks. 
The boundary CTTS/WTTS is set represented by the horizontal dashed line at EW 10\AA.
The probability of detecting H$_2$ in CTTS seems to be correlated with accretion
signatures (H$\alpha$, $UV$ excess).
In WTTS H$_2$ emission is observed preferentially towards sources with high 
X-ray luminosity. Figure taken from Bary et al. (2008). }
\label{fig:6}
\end{figure}

In classical T Tauri stars (CTTS) H$_2$ near-IR lines are preferentially observed in objects 
that exhibit signatures of active accretion, 
that is, 
objects with large H$\alpha$ equivalent widths and strong UV excess 
(Carmona et al. 2007, Bary et al. 2008, see Figure 6).
In CTTS there is no apparent correlation between the X-ray luminosity, disk mass
and the presence of the near-IR H$_2$ lines (Carmona et al. 2007).
The H$_2$ 1-0 S(1) line has been detected in a few WTTS 
(Bary et al. 2002, Bary et al. 2008, see Figure 5).
This indicates that a fraction of such objects still have a gaseous disk. 
The WTTS in which the H$_2$ 1-0 S(1) line has been detected are sources
exhibiting large X-ray luminosity (Bary et al. 2002, Bary et al. 2008, see Figure 6).

H$_2$ near-IR lines have been observed in relatively old objects such as ECHAJ08843.3-7905 in the
$\eta$ Chamaleontis cluster (6 Myr, Ramsay Howat \& Greaves 2007) 
and TW Hya (Bary et al. 2000) in the TW Hya association (8-10 Myr).
H$_2$ lines have been detected in transitional disks in Chamaleon (Bary et al. 2008),
and in the close binary star GV Tau (Doppmann et al. 2008).
In the transitional disks in Chameleon in which H$_2$ near-IR emission has been observed,
[Ne II] emission at 12 $\mu$m has been also detected by Spitzer (Bary et al. 2008). 
This suggests a common excitation mechanism for both lines (see Section 3.7). 

The main strength of the H$_2$ lines in the near-IR lies in their ability to probe very small 
quantities of hot gas in the inner disk if the right excitation conditions are present
(accretion, UV photons, X-rays).
This strength can be exploited to search for gas in environments where other
gas diagnostics give negative results (i.e., WTTS).
In a similar way, non-detections allow to set stringent upper limits on the presence of hot gas
in the inner disk (e.g., Carmona et al. 2007 found no evidence of H$_2$ emission in the debris disk
49 Cet, and constrained the mass in the inner most disk to be lower than a few lunar masses).
Observationally, H$_2$ lines in the near-IR have the advantage that they are observed in the
K band. The transmission of the atmosphere in the K band is relatively good; 
hence, H$_2$ lines in the near-IR can be searched in relatively faint objects with large telescopes. 
There is potential for perfoming large surveys for H$_2$ emission in the near-IR.

The main disadvantage  of H$_2$ near-IR emission is that it is also 
produced by shocks in outflows. 
Thus, to reduce the risk of confusion with shocks,
high spectral resolution observations with precision higher than 5 km/s and 
observations at high angular resolution with AO are needed.
For the interested reader, 
Beck et al. (2008) and Gustafsson et al. (2008) present examples of H$_2$ integral
field spectroscopy obtained with AO in T Tauri stars.
In the sources observed by those autors, 
we can observe the contribution of outflow emission to the H$_2$ 1-0 S(1) observed.
Finally, note that near-IR H$_2$ only trace the hot
H$_2$ in the innermost disk (R$<$few AU). Consequently, an additional limitation
is that H$_2$ lines in the near-IR can {\it not} be used to derive the total disk mass
or the gas mass in the giant planet forming region of the disk
(note that most of the mass of the disk is in cold gas at R$>$20 AU). 

\subsubsection{H$_2$ pure-rotational emission in the mid-IR}
H$_2$ pure-rotational emission in the mid-IR traces warm gas at temperatures
generally from 150 K up to 1000 K.
Thus, H$_2$ mid-IR lines have the unique potential of tracing the gas in the
giant planet formation region of the disk at R $>$ 2 AU.
With present sensitivities (5$\times 10^{-15}$ erg/s/cm$^2$) 
a few Earth masses of warm gas in sources at 140 pc can be probed.
H$_2$ mid-IR emission arises from transitions 
between the rotational levels ($\Delta$J=2) of the ground vibrational state ($\upsilon$=0).
Typical lines searched are the $\upsilon$=0-0 S(0) line (J=2-0) at 28 $\mu$m (visible only from space),
the 0-0 S(1) line at 17 $\mu$m, the 0-0 S(2) line at 12 $\mu$m and the 
0-0 S(3) line at 9.6 $\mu$m.

Following the initial detection claims of H$_2$ 0-0 S(0) 
and 0-0 S(1)  emission from observations with ISO\footnote{Infrared Space Observatory} 
(Thi et al. 2001),
several ground-based (Richter et al. 2002; Sheret et al. 2003;  Sako et al. 2005,
Martin-Za\"idi et al. 2007 \& 2008b; Bitner et al. 2007 \& 2008; 
Carmona et al. 2008b) 
and space (Hollenbach et al. 2005; Pascucci et al. 2006; Lahuis et al. 2007)
searches for H$_2$ mid-IR emission have been performed.
ISO detections remain controversial, 
as all subsequent efforts do not confirm the high H$_2$ emission fluxes reported by Thi et al.

At the time of writing, positive detections of  H$_2$ mid-IR emission 
have been reported only in a handful of objects:
the Herbig Ae/Be stars AB Aur (Bitner et al. 2007) and
HD 97048 (Martin-Za\"idi et al. 2007), and the T Tauri stars Sz 102, EC 74, EC 82, Ced\_110\_IRS6,
EC92, ISO-Cha237 (Lahuis et al. 2007), DoAr 21, Elias 29, GSS 30. GV Tau N and Hl Tau 
(Bitner et al. 2008). 
In total, H$_2$ pure-rotational emission has been detected only in 13 young stars with disks from $\sim$ 114 objects searched 
(here we exclude 14 ISO H$_2$ detections in 18 stars observed, and the searches in 16 optically thin disks
of Hollenbach et al. 2005 and Pascucci et al. 2006).  
The main result of these searches are: (i)  H$_2$ mid-IR emission is relatively rare ($\sim11\%$ of the sources),
and (ii)  in the sources where the emission is detected, the gas is heated by an additional mechanism.

The numerous non detections can be understood in a first approximation
under the frame of the Chiang \& Goldreich (1997) two-layer model
of an optically thick disk.
In this model, 
only emission from the surface molecular layer of the disk is observed (the mid-plane is optically thick).
Since the amount of gas in the surface layer is very small,
if the dust and gas are at equal temperatures and the gas-to-ratio is equal to 100,
then the expected thermal emission of H$_2$ from the surface layer is too weak to be detected
($\sim$10$^{-16}$-10$^{-17}$ erg/s/cm$^2$, Carmona et al. 2008).
In the context of the two-layer model,
the few detections can be explained
if the temperature of the gas is allowed to be at twice the temperature of the dust and
if the gas-to-dust ratio is larger than 1000 (Carmona et al. 2008).
This can occur under certain physical circumstances such as a high UV or X-ray radiation fields or
if dust coagulates and sediments towards the mid-plane of the disk. 

The two-layer model is just an approximation to the real structure of the disk;
however, similar conclusions have been reached from more detailed modeling.
Sophisticated models of H$_2$ emission from disks (Nomura et al. 2005, 2007) assuming typical 
X-ray and UV-fluxes from pre-main sequence stars predict flux levels below present detection limits 
(both in line flux and in the line contrast with the continuum).
Models from optically thick disks with enhanced levels of  UV and X-ray radiation (e.g., Gorti \& Hollenbach 2008) 
can account for the H$_2$ fluxes reported.
Finally, note that H$_2$ mid-IR emission has been reported up to the 0-0 S(9) line 
(indication that an extra source of heating is present) 
and that it has been observed in a few transitional disks (e.g., Sz 102) and in the WTTS DoAr 21 (Bitner et al. 2008).

In summary, H$_2$ mid-IR lines have the potential of tracing the gas in the giant planet forming region
of the disk, but their main limitation is that they are too weak to be detected in a large number
of sources. In addition, if the disks observed are optically thick, then mid-IR H$_2$ lines  
only trace the small amount of gas in the surface layer of the disk. 
Therefore, H$_2$ mid-IR lines in optically thick disks cannot be used to derive constraints in the total disk mass.
In the objects in which we are able to detect the emission, 
H$_2$ appears  be heated by an additional mechanism. 
In this case the LTE approximation is no longer valid and the H$_2$ masses derived
can be overestimated. 

H$_2$ mid-IR emission studies from the ground are so far limited
to relatively bright sources with fluxes typically larger than 1 Jy.
Future mid-IR infrared instrumentation such as high-resolution mid-IR spectrograph
EXES in the airplane observatory SOFIA, and the mid-IR 
spectrograph MIRI in the James-Webb space telescope will  increase the sensitivity and 
will allow the study of H$_2$ mid-IR emission in 
larger number of objects, in particular in CTTS.  
   
\subsubsection{H$_2$ electronic transitions in the UV}
H$_2$ electronic transitions in the UV probe two kinds of gas:
(i) in emission when the disk is seen face on, they trace highly excited
gas at temperatures of a few thousand Kelvin (R$<<$ 1 AU).
(ii) in absorption when the disk is seen close to edge-on, they probe
colder gas at few hundred K.  
H$_2$ UV lines are observed in absorption against the bright continuum in hot stars  
or against broad emission lines (e.g., OVI at $\lambda$ 1032-1038\AA) in cool stars. 
H$_2$ electronic transitions occur between the vibrational levels of the first $(B)$ 
or the second $(C)$ electronic states and the ground 
vibrational level of the ground electronic state ($X$).
Given the low transmission of the atmosphere at UV-wavelengths,
H$_2$ UV transitions are studied primarily with space instrumentation 
(e.g., IUE, HST, FUSE)\footnote{IUE : International Ultraviolet Explorer; 
HST: Hubble Space Telescope; FUSE: Far Ultraviolet Spectroscopic Explorer.}.

H$_2$ lines are excited by the far-UV (FUV) flux of the central star,
or by shocked gas in outflows.
Warm H$_2$ gas can absorb photons all over the FUV;
consequently, a flat radiation field would excite a very 
large number of densely packed H$_2$ lines and a pseudocontinuum
is produced.
In the case of CTTS, the FUV flux is dominated by emission lines,
in particular by Ly$\alpha$. Hence, in CTTS, we observe principally
Ly$\alpha$-band (B-X) transitions. Nevertheless, 
Werner-band transitions (C-X) are observed as well.

H$_2$ electronic transitions have been searched towards T Tauri stars
(Valenti et al. 2000; Errico et al. 2001; Ardila et al. 2002; Wilkinson et al. 2002; 
Herczeg et al. 2002, 2004, 2005 \& 2006; Walter et al. 2003; Bergin et al. 2004),
Herbig Ae/Be stars (Roberge et al. 2001; Lecavelier des Etangs et al. 2003; 
Bouret et al. 2003; Grady et al. 2005; Martin-Za\"idi et al. 2004, 2005 \& 2008a),
accreting Brown Dwarfs (Gizis et al. 2005) 
and in close to edge-on debris disks ($\beta$ Pic, Lecavelier des Etangs et al. 2001,
Martin-Za\"idi et al. 2008; AU Mic, Roberge et al. 2005, France et al. 2007).

In the case of T Tauri stars, H$_2$  UV lines have been detected in
several CTTS and in at least one WTTS (V836 Tau). 
Nontheless, a large fraction of the H$_2$ detections in CTTS appears to be emission from outflows.
The observed lines are blueshifted (e.g., Ardila et al. 2002) 
and/or spatially extended (e.g., Herczeg et al. 2006).  
So far, only in very few cases quiescent spatially unresolved  H$_2$ emission in the UV 
has been reported. The most notably example is the T Tauri star TW Hya.
In this source, the lines observed are excited by Ly$\alpha$ fluorescence. The
emission is consistent with emission from the surface of a hot inner disk (Herczeg et al. 2002)

In the case of Herbig Ae/Be stars, H$_2$ UV lines have been studied mostly in absorption
at 1032-1038 \AA. Evidence for different origin/excitation 
for the lines is observed as function of the stellar mass.
For the massive Herbig Be stars (spectral types B2 to B8), 
the H$_2$ lines are consistent with a photodissociation region (PDR) in large circumstellar envelopes
(Martin-Za\"idi 2008a).  
For Herbig Ae stars, the line of sight generally appears {\it not} to pass through the disk. 
The H$_2$ lines observed are consistent with warm gas in the circumstellar environment of the stars 
(e.g., an envelope in AB Aur, Roberge et al. 2001) or cold gas in the line of sight (e.g., HD 141569 Martin-Za\"idi et al. 2005).
In Herbig Ae stars, the H$_2$ absorption lines are not consistent with PDR models (Martin-Za\"idi 2008a).
In some cases, if the inclination and flaring of the disk are adequate,
we may see a thin layer of a flared circumstellar disk (e.g., HD 100456 \& HD 163296,
Lecavelier des Etangs et al. 2003). 

In the case of the debris disk $\beta$ Pic, no strong H$_2$ lines either in emission or absorption have been 
detected. Only a marginal detection of the J=0 H$_2$ line was reported (Lecavelier des Etangs et al. 2001).
This was interpreted as evidence of lack of gas in the  $\beta$ Pic disk. 
In the case of AU Mic H$_2$ was not detected in absorption, but it was detected in emission.
The estimated gas mass was of $10^{-4}-10^{-6}M_{\rm earth}$ (France et al. 2007).

The main advantage of the H$_2$ lines in the UV are:
(i)  if observed in emission, they can  probe small amount of very hot gas (T$\sim$2500 K) in the innermost part of the disk;
(ii) if observed in absorption, they can trace small amounts of colder  (T$\sim$ 50-hundreds of K) gas.
H$_2$ lines in the UV have allowed us to unveil the presence of hot gas in the inner disk
in WTTS and transitional disks. 

Their main limitations are: 
(i) in most of the CTTS, the H$_2$ emission in the UV originates in outflows; 
(ii) in most of the Herbig Ae stars where H$_2$ has been observed in absorption,
the probed gas is not in the disk, but instead in circumstellar ambient material
or cold gas along the line of sight;
(iii) absorption studies probe a narrow pencil-beam that limits its application in
sources observed far from edge-on;
(iv) in the case of detections in emission, 
the interpretation of the results (i.e., column densities) requires complex
modeling based on the poorly constrained  FUV field of the central star.
Finally, uncertain conversion factors are needed to extrapolate the mass
of gas probed to the total amount of gas in the disk.
   
\subsection{[OI] emission at 6300 \AA}
[OI] emission at 6300 \AA~ is a tracer of atomic gas in the surface layers of flared disks. 
The line traces up to a 100 AU depending on the geometry of the disk. 
[OI] forbidden emission at 6300 \AA~ is a diagnostic that started to be explored recently.
It has been studied  
in a growing number of Herbig Ae/Be stars with high-resolution optical spectroscopy
(Acke \& van den Ancker 2005 \& 2006; van der Plas et al. 2008a; Fedele et al. 2008).
The majority of the studied sources display narrow (width $<$ 50 km/s) single peaked profiles;
however, several objects exhibit double-peaked profiles.
In such objects, 
the low velocities with respect to the star's rest velocity, 
the symmetry of the features and the peak-to-peak separation
are consistent with line emission from a circumstellar disk in Keplerian rotation. 
                  
The strength of the [OI] line appears to be strongly correlated with the SED 
shape of the sources (i.e., disk geometry).
Following the SED classification scheme of Meeus et al. (2001),
Acke et al. 2005 found that    
Herbig Ae/Be from group I (flared disks) exhibit a [OI] emission  
stronger than those of group II (self-shadowed disks).
In addition, the feature was found to be absent in an important fraction (40\%) 
of group II sources.
Acke et al. found also that the [OI] luminosity is correlated to the PAH luminosity of
the sources. 
Acke et al. proposed that the [OI] line arises from the photodisociation by UV radiation  
of OH and H$_2$O molecules in the surface layer of flared disks.
The conclusions shown by Acke et al. require fractional OH abundance $\sim \epsilon$(OH) ~10$^{-7}$-10$^{-8}$.

Based on the analysis of the [OI] emission line
profile in HD 100546, Acke et al. (2006) 
found evidence of a gap present at 10 AU in HD 100546's disk.
They suggested that such a gap was likely induced by a planetary-mass companion
of 20 M$_J$ located at 6.5 AU.  
Temporal changes in the [OI] line profile were reported as well.
Acke et al. concluded that such changes are related to  inhomogeneities 
in the [OI] disk emitting region.

By analyzing the [OI] line, van der Plas et al. (2008a) found evidence 
of the existence of a puffed-up 
inner rim followed  by a shadow in HD 101412.
Fedele et al. (2008) employed a combined analysis of the [OI] emission 
and mid-IR interferometry observations found evidence of different
dust and gas vertical structure inside 2 AU in the 
disks of HD 101412 and HD 135344B. 
Fedele et al. interpreted their findings as the result of 
dust/gas decoupling in the inner disk. 
They suggested that they may exist an evolutionary sequence
from flared disks to flat disks due to the combined action of gas-dust decoupling, grain growth,
and dust settling.

[OI] emission at 6300 \AA~ has the advantange of being observable in the optical;
therefore, statistical studies of large numbers of objects are possible 
at high-spectral resolution in the future.
One limitation of the diagnostic is that [OI] emission  can 
produced by shocked gas in an outflow. 
Thus high spatial and spectral resolution
observations are required to avoid confusion with emission from outflows.

\subsection{[Ne II] emission at 12.8 $\mu$m}
Recently, Pascucci et al. (2007) and Lahuis et al. (2007) reported the detection of  
[Ne II] 12.8 $\mu$m emission from T Tauri stars with Spitzer. 
Pascucci et al. (2007) detected the line in four out of six targets (all with faint
mid-IR continuum with respect to classical  T-Tauri stars of the same spectral type) 
and Lahuis et al (2007) in 15 out of 76 T Tauri stars.  
These first detections were followed by ground-based detections in TW Hya by Herczeg et al. (2007).
The [Ne II] line has been observed towards optically thick disks, as well as 
in the transitional disk GM Aur and the circumbinary disk
CS Cha (Espaillat et al. 2007).

Given that neon cannot be ionized by photons with energies of less than 21.4eV,
the detection of [Ne II] emission from disks attracted the attention of the disk community.
The [Ne II] line could provide evidence of higher energy photons irradiating the disk,
in particular extreme-UV (EUV) photons (Gorti \& Hollenbach, 2008; Alexander et al. 2008) 
or X-rays (Glassgold et al. 2007, Meijerink et al. 2008).
Nonetheless, [Ne II] emission can be produced by shocked gas in outflows (e.g., Hollenbach \& McKee 1989).
At the present, it is not well constrained which of these scenarios is taking place.
Spitzer's spatial and spectral resolution are too modest to distinguish between the
different scenarios proposed.

Models of X-ray irradiated disks by Glassgold et al. (2007) 
predicted that the ionization fraction of neon
is proportional to the square root of the X-ray luminosity.
However, models by Meijerink et al. (2008) rather found that there is
a linear proportionality between the [NeII] line luminosity and the X-ray luminosity.
Pascucci et al (2007) reported a tentative correlation between the [Ne II] and the
X-ray luminosities.
However, only about 30\% of the sources with [Ne II]
detections from Lahuis et al. (2007) are identified as X-ray sources
(either due to the lack of X-ray emission, sensitivity limited X-ray searches, or object geometry).

In the UV model of Gorti \& Hollenbach (2008), 
EUV photons from the stellar chromosphere and/or from accretion create a region similar to a HII
region at the surface of the disk. 
The EUV are absorbed in the inner 10 AU of the disk, 
and a EUV photon luminosity $\sim$10$^{41}$ erg/s would be able to produce
detectable [NeII] line fluxes.
But, in the EUV model, a stellar wind greater than $\sim10^{-10} M_\odot/{\rm yr}$
could drastically reduce the UV-photons reaching the disk. Stellar winds
of this magnitude correspond to typical CTTS accretion rates. 
Thus, lower [Ne II] luminosities are expected for CTTS stars simple 
due to the accretion and the related wind rates. 
Nonetheless,  several detection from Lahuis et al. (2007) are from CTTS.

In an alternative model, Alexander et al. (2008) suggested that the [Ne II]
line is produced in a photoevaporating wind. 
Alexander et al. predicted broad (30-40 km/s) double peaked [Ne II] lines 
when the stars are viewed close to edge-on and 
narrower ($\sim$10 km/s), slightly blue-shifted lines when viewed face-on.
High spatial and spectral resolution is required for testing this model. 

In an additional scenario, high-velocity shocks can entrain ionized lines (e.g., Hollenbach \& McKee 1989). 
Van den Ancker et al. (1999) detected [Ne II] emission in the ISO spectra of the T Tau triplet. 
This system contains three stars of which at least one is a strong X-ray source, 
as well as regions of shocked gas in the immediate vicinity. 
Due to the large beam (27" $\times$ 15") it was not clear from where in the system the emission comes. 
Van den Ancker et al. proposed that the emission arose in J-type (dissociative) shocks resulting from 
the interaction of the outflow of T-Tau south with ambient material. 
T~Tau exhibits a very high $L_{\rm{[Ne\,II]}}/L_{\rm{X}}$ ratio, 
hinting that processes other than X-ray irradiation are important.
Recent high spectral and spatial resolution observations of the [Ne~II] line in T Tau with the VLT
spatially resolved the various components of the system (van Boekel et al. in prep.).
Van Boekel et al. found that the vast majority of the [Ne~II] flux appeared to be 
associated with an outflow from T~Tau~S, 
and that only a small fraction ($\sim$5$-$10\%) of the [Ne~II] emission was
directly related to the X-ray bright Northern component. 
van Boekel et al. observations showed that if strong accretion/outflow activity is present, 
shocks were the main mechanism for producing [Ne~II] emission in T Tau.

Finally, note that in three young stars that exhibit the {\it Spitzer} [Ne~II] line,
the H$_2$ 1-0 S(1) line at 2.12 $\mu$m~
was detected as well:
TW Hya (Bary et al. 2003), GM Aur (Shukla et al. 2003) and CS Cha (Bary et al. 2008).
Bary et al. (2008) pointed out that 
simultaneous presence of the [Ne~II] line  and  
{\it quiescent} H$_2$ disk emission suggests a shared excitation mechanism of the
gas in the disk by the central star's high-energy photons.
The same high-energy photons that stimulate
the [Ne II] emission could also produce the {\it quiescent} H$_2$ emission observed.
However, as previously discussed,
H$_2$ near-IR emission can be also produced by shocks.
In such a case, objects displaying the [Ne~II] lines,
may show the H$_2$ near-IR emission  velocity shifted or extended. 
In such objects, H$_2$ and [Ne II] emission will be produced by shocks.
Spectrally and spatially resolved observations are required to test both scenarios.

In summary, [NeII] emission has the potential of unveiling 
the effects of X-ray or UV-ray in the surfaces of disks.
But, since [NeII] emission can be produced by shocks, 
it is necessary to perform observations with higher spatial and spectral
resolution to determine the dominant excitation mechanism.
In addition, observations of larger samples are
required to establish statistically meaningful correlations between the
presence/strength of [NeII] emission and physical properties of the 
central star (i.e., spectral type, accretion rate, age, X-ray and UV-luminosity).

\subsection{H$_2$O, C$_2$H$_2$, HCN, OH molecular emission in the near and mid-IR} 
How water is transported to the surface of habitable planets is one
fundamental and fascinating question in planet formation theory.
Water and other simple organic molecules are expected to be
abundant -in gas phase- in the inner regions of the disk (R$<$5 AU).
Nevertheless, observational measurements of water in the planet forming region of 
disks are relatively recent.
First detections of hot water vapor were reported 
by Carr et al. (2004) and Thi \& Bik (2005b) in the 
near-IR spectrum of young stellar objects 
exhibiting the 2.3 $\mu$m CO bandhead emission (e.g., SVS 13, DG Tau and 51 Oph).

The observed H$_2$O displays a characteristic excitation temperature of 
$\sim$1500 K. Since this temperature is cooler than 
the temperature of the CO overtone emission (2500 K)
the observed water emission should be produced in a region exterior to the CO 2.3 $\mu$m
emitting region. Simultaneous modeling of the CO and steam emission revealed that the 
H$_2$O/CO ratio is lower to that expected in chemical equilibrium (Carr et al. 2004).
Physical processes other than equilibrium chemistry should be responsible of the 
mesaured H$_2$/CO abundances. Thi \& Bik (2005) reproduced the H$_2$/CO abundance observed
in 51 Oph with gas at 1600 K and an enhanced UV field over gas density ratio.

These first detections of H$_2$O were followed by detections in the IR 
of simple molecules such as C$_2$H$_2$ and HCN. 
They were observed in absorption in the near-IR spectrum of the T Tauri binary
GV Tau N (Gibb et al. 2007 \& 2008) and in absorption in the mid-IR spectrum of 
the low-mass young 
stellar object IRS 46 (Lahuis et al. 2006).
The temperaures measured were of a few hundred K. 
The lines originate most likely in the inner disk (R$<$6 AU) of the sources.
However, there is also the possibility that they originate in a disk wind.

The high sensitivity reached with IR spectroscopy from space and ground-based
high spectral resolution allowed the detection of emission of
simple organic 
molecules from the terrestrial planet region of the disk (R$<$5 AU). 
Carr \& Najita (2008) reported the detection of HCN, C$_2$H$_2$, CO$_2$ 
water vapor and OH emission in the Spitzer's mid-IR spectrum of the T Tauri star AA Tau.
Modeling of the observed profiles indicates that the observed gas has
temperatures from 500 to 900 K and is located within 0.5 and 3 AU.
One important finding is that the abundances relative to CO found 
are much higher than those obtained from models of hot molecular cores, 
this suggests that substantial molecular synthesis occurs within the disk.

Salyk et al. (2008) reported the detection of H$_2$O emission at 3 $\mu$m and 10-20 $\mu$m
and OH emission at 3 $\mu$m in the T Tauri stars AS 205a and DR Tau. 
They measured excitation temperatures of 1000 K. 
Based on the velocity wings of the lines, they concluded that the observed emission arises up 
to radius of 1 AU. 

Mandell et al. (2008) found OH emission at 3 $\mu$m in the Herbig Ae stars 
AB Aur and MWC 758. Optically thin LTE models of the observed lines
revealed excitation temperatures $\sim$700 K and an OH emitting region extending
up to 1 AU. The observed OH lines can be explained by collisional excitation models,
and in the case of AB Aur by a UV fluorescence model too.  

Pascucci et al. (in prep.) detected with Spitzer
C$_2$H$_2$ and HCN at 7-14 $\mu$m in a large number of young stellar objects with disks
They found that the abundance of these simple organic molecules
appears to be different between stars cooler than the Sun and sun-like stars.

The study of simple organic molecules in the near and mid-IR is growing. 
The main advantage of these diagnostics are:
(i)  they trace the gas in the terrestrial planet forming region of the disks, and 
(ii) they have the potential of constraining the chemical processes in this region. 
From the observational point of view, an additional 
advantage is that within a single observation set-up, multiple 
tracers are observable simultaneously. 
In addition, the fact that some of  the lines are observable from the ground at 3 $\mu$m
open the door to the study of a large number of objects at high spectral and spatial resolution.

One limitation of these gas tracers is that the interpretation of the observations 
generally requires complex modeling that involves: (i) assumptions in the structure of the 
inner disk; (ii)  poorly constrained physical quantities (e.g., the UV field, inclination of the inner disk);
and (iii) poorly constrained physical processes (e.g., excitation mechanisms, chemistry, radial and vertical mixing in the disk). 
On the other hand, these limitations can become an advantage, because,
even if observations can be interpreted with different models, observations
can also rule out scenarios that are not consistent with the measurements.
It is interesting that future observations of simple molecules in the inner disk
could reveal a variety of disk chemistry until now unsuspected.
Finally,  note that in very short time Spitzer will not offer mid-IR spectroscopy anymore.
However, future mid-IR spectrograph MIRI planned for JWST
will be an ideal tool for this research.

\section{Conclusion}
In this contribution I have discussed several observational diagnostics that
allow to study the gas in protoplanetary disks highlighting
the strengths and limitation of each diagnostic. Here, I summarize 
important general points that one should bear in mind 
about observational constraints of gas in disks.
\begin{itemize}
\item Disks have a radial temperature structure; therefore, different 
gas diagnostics are employed for probing different regions of the disk.
UV probes the hottest gas in the innermost disk (R$<<$1 AU, T$>$2000 K), 
NIR and MIR probes hot and warm  
gas in the inner disk (R$\sim$0.1-20 AU, T$\sim$100-2000 K), 
(sub)-mm probes cold gas in the outer disk (R$>20$ AU, T$\sim$20-100 K). 
Note that disks have typical sizes of several hundreds of AU.
\item To observe a line from gas (in emission or absorption) the dust in the medium where the
line is produced must be optically thin at the observed wavelength, 
or the gas should be at a higher temperature than the dust. 
\item An optically thick disk has a vertical temperature structure. 
The surface layer is hotter than the interior mid-plane layer.
The dust in the surface layer is optically thin, the dust in the interior layer (of the inner disk) is optically thick.
In optically thick disks, hot and warm gas emission lines from the UV to the IR are produced in the surface layer.   
Note that the amount of mass in the surface layer is much smaller than the amount of mass in the interior layer. 
Gas lines in the IR probe only a limited amount of material.   
H$_2$ lines in the IR are optically thin. CO lines in the IR can be optically thick.
\item Sub-mm lines trace only the cold outer regions of the disk at R$>$20 AU. Planets
usually are not expected to form at this distances. Still, most of the mass of the disk
is at R$>$20 AU. 
\item CO in the (sub)-mm is not a reliable disk gas mass tracer because (i) CO freezes onto
dust grain surfaces, (ii) the CO/H$_2$ conversion factor is unknown in disks (iii) the emission
is likely optically thick.
\item Be aware that so far we do not measure directly the mass of gas in disks. We deduce
the disk mass from dust continuum emission in the (sub)-mm from cold dust in the outer disk
(R$>$20 AU). The deduced mass depends on the dust opacity, dust temperature and the 
dust-to-gas ratio assumed. 
The disk mass observed is from material located at R$>$50 AU,
and disk evolutionary models are typically of disks of R$<$50 AU.
The mass and surface density profiles of gas at R$<$20 AU are poorly constrained
from observations.
\item Modeling is required to interprete line observations.
Conventional assumptions are that the observed gas is at LTE and is optically thin.
The rotational diagrams employed for deducing the temperature of the 
observed gas make use those two assumptions.
To deduce the inner most radius of the observed emission (i.e., R$_{\rm in}$) from
line profile modeling, it is required to assume the disk inclination and 
the dependence of the intensity as a function of radius ($I(R)$). 
Inclinations are typically deduced from (sub)-mm imaging and extended to the inner disk.
On the other hand, if we want to deduce independently the inclination (e.g., to search for
evidence of warps) we need to assume (or calculate) R$_{\rm in}$ and $I(R)$.
\item Most of the gas tracers (e.g., H$_2$, [Ne II]) can be produced by shocked emission from
outflows as well.
High spectral/spatial resolution is required for determine whether the emission is
observed at the velocity of the central star and whether it is spatially extended or not.
\item Gas emission could be excited by collisions with dust grains, shocks and UV or X-rays.
Line ratios are generally employed to discriminate between the different excitation
mechanisms.
\item Gas could be present in a disk even when dust emission is weak or not present. 
Gas emission lines have been observed in the inner regions of transitional disks and
in several WTTS. 
\item The study of  gas in disks acquired momentum with the advent of high-resolution
ground IR spectrographs and high-sensitivity IR spectrographs in space. 
\item A new chapter in gas studies was recently written with the detection of emission of simple molecules
 in the inner disks. 
\end{itemize}

At the beginning of the paper I highlighted some fundamental science questions that required
direct observational constraints from the gas in the disk. 
To conclude, I would like to address them again from the actual state of the observational study of gas.

{\it (i) How long does the protoplanetary disk last? }
In general terms, disk gas diagnostics are not detected when the signatures of the dust in the disk
have disappeared. Thus, as a first approximation the gas in the disk disappears 
almost simultaneously with the signatures from small dust particles. Nonetheless, we should be aware of a few details:
(a) with the exception of H$\alpha$ emission, there are many sources for which we know
that they have a disk, but given sensitivity issues, we are not able to detect any gas tracer; 
(b) gas has been observed in several WTTS stars (e.g., H$_2$ in the UV and NIR),
    and in transitional disks (e.g., H$_2$ in the near-IR, CO at 4.7 $\mu$m, [Ne II] at 12 $\mu$m);
(c) unbiased surveys for hot/warm gas in a large sample of WTTS have yet not been performed;
(d) surveys of H$\alpha$ at high spectral resolution need to be done in larger samples of WTTS
in star-forming regions of different ages. If the accretion rate and spectral resolution 
are low, one can easily miss objects accreting gas at low accretion rates.
In summary, an estimate of disk gas lifetime independent of dust is still required.
   
{\it (ii) How much material is available for forming giant planets? }
On this aspect our hopes are fainter. Several surveys for H$_2$ emission
in the MIR showed that the emission is observed only in a handful of objects,
and that in these objects the gas is heated by an additional (to dust collisions)
heating mechanism (UV, X-ray excitation). Although these results confirm our 
two-layer picture of the disk, in practice these results mean that we are blind to the 
disk's mid-plane and that we are unable to know directly how much mass is in the giant 
planet forming region of the disk. The good news are two fold: (a) in objects with
data at several wavelengths, an educated guess can be found from modeling the surface density (see next point);
(b) at some point the disk should become optically thin -by dust evolution- 
then we will be able to measure the gas mass.

{\it (iii) How do the density and temperature of the disk vary as a function of the radius? }
There is a growing number of sources (e.g., TW Hya, AB Aur) 
for which we start to have information
from several gas tracers at wavelengths from the UV to the (sub)-mm.
In such sources, we can attempt to constrain the disk structure employing a
disk model aimed to fit all the available data.
Therefore, at least for a handful of objects, by modeling multiwavelength spectroscopy,  
we have the potential to constrain the density and temperature as a function of radius.
However, we should always keep in mind that we observe emission from the disk's surface layer
and that the mid-plane temperature and density will be deduced from the 
model. Hence, the final gas mass deduced will be model dependent.  

{\it (iv) What are the dynamics of the disk? }  
In this aspect we had good news. To detect gas lines in the IR we commonly use the highest spectral
resolution available. At the present, the resolution is sufficiently high to spectrally resolve 
the lines. This allows the modeling of the line shape; therefore, constraints
in the dynamics of the emitting gas can be derived. 
Recently, with the combination of AO and high-resolution spectroscopy, 
it has been possible to spatially resolve gas disk emission directly 
or to spatially resolve the spectroastrometric signal with spectra taken at different slit orientations. 
So far the observed gas is consistent with gas in Keplerian rotation. 
This kind of studies should be extended in the future to a larger number of systems.

\section*{Acknowledgments}
I would like to thank the organizers of the Ascona planet formation meeting for
inviting me to give the observations of gas in disks talk, 
and to R. Nelson, C.P. Dullemond, C. Clarke and S. Fromang for kindly 
accepting that I present my talk as in individual contribution in this volume.
I am grateful for the detailed suggestions in the manuscript provided by M. Audard, 
C. Baldovin-Saavedra and F. Fontani.
I would like to warmly thank M. van den Ancker, Y. Pavluychenkov, 
M. Goto, C.P. Dullemond, Th. Henning, C. Martin-Za\"idi, G. van der Plas and
D. Fedele for all the discussions about gas in disks during the last years.
Part of this work is supported by a Swiss National Science Fundation grant to M. Audard.


\begin{thebibliography}{}
%
%
\bibitem{Acke et al. (2005)}  Acke, B., van den Ancker, M.~E., \& Dullemond, C.~P.,  {\it "[O I] 6300 \AA; emission in Herbig Ae/Be systems: Signature of Keplerian rotation"},  Astronomy and Astrophysics,  436,  209,  2005

\bibitem{Acke and van den Ancker(2006)}  Acke, B., \& van den Ancker, M.~E.,  {\it "Resolving the disk rotation of HD 97048 and HD 100546 in the [O I] 6300 \AA; line: evidence of a giant planet orbiting HD 100546"},  Astronomy and Astrophysics,  449,  267, 2006 

\bibitem{Akeson et al.(2005a)}  Akeson, R.~L., et al.,  {\it "Keck 
Interferometer Observations of Classical and Weak-line T Tauri Stars"},  
Astrophysical Journal,  635,  1173,  2005a 

\bibitem{Akeson et al.(2005b)}  Akeson, R.~L., et al.,  {\it "Observations 
and Modeling of the Inner Disk Region of T Tauri Stars"},  Astrophysical 
Journal,  622,  440,  2005b

\bibitem{Alexander(2008)}  Alexander, R.~D.,  {\it "[NeII] emission-line 
profiles from photoevaporative disc winds"},  Monthly Notices of the Royal 
Astronomical Society,  391,  L64,  2008 

\bibitem{Andrews and Williams(2005)}  Andrews, S.~M., \& Williams, J.~P.,  {\it "Circumstellar Dust Disks in Taurus-Auriga: The Submillimeter Perspective"},  Astrophysical Journal,  631,  1134,  2005 

\bibitem{Ardila et al.(2002)}  Ardila, D.~R., Basri, G., Walter, F.~M., 
Valenti, J.~A., 
\& Johns-Krull, C.~M.,  {\it "Observations of T Tauri Stars using Hubble Space Telescope GHRS. I. Far-Ultraviolet Emission Lines"},  Astrophysical Journal,  566,  1100,  2002 

\bibitem{Bary et al. (2002)}  Bary, J.~S., Weintraub, D.~A., \& Kastner, J.~H.,  {\it "Detection of Molecular Hydrogen Orbiting a ``Naked'' T Tauri Star"},  Astrophysical Journal,  576,  L73,  2002

\bibitem{Bary et al. (2003)}  Bary, J.~S., Weintraub, D.~A., \& Kastner, J.~H.,  {\it "Detections of Rovibrational H$_2$ Emission from the Disks of T Tauri Stars"},  Astrophysical Journal,  586,  1136,  2003 

\bibitem{Bary et al.(2008)}  Bary, J.~S., Weintraub, D.~A., Shukla, S.~J., 
Leisenring, J.~M., \& Kastner, J.~H.,  {\it "Quiescent H$_2$ Emission From Pre-Main-Sequence Stars in Chamaeleon I"},  Astrophysical Journal,  678,  1088,  2008

\bibitem{Beck et al. (2008)}  Beck, T.~L., McGregor, P.~J., Takami, M., \& Pyo, T.-S.,  {\it "Spatially Resolved Molecular Hydrogen Emission in the Inner 200 AU Environments of Classical T Tauri Stars"},  Astrophysical Journal,  676,  472,  2008 

\bibitem{Beckwith et al. (1978)}  Beckwith, S., Gatley, I., Matthews, K., \& Neugebauer, G.,  {\it "Molecular hydrogen emission from T Tauri stars"},  Astrophysical Journal,  223,  L41,  1978

\bibitem{Beckwith and Sargent(1993)}  Beckwith, S.~V.~W., \& Sargent, A.~I.,  {\it "Molecular line emission from circumstellar disks"},  Astrophysical Journal,  402,  280,  1993 

\bibitem{Bergin et al.(2004)}  Bergin, E., et al.,  {\it "A New Probe of 
the Planet-forming Region in T Tauri Disks"},  Astrophysical Journal,  614,  
L133,  2004 

\bibitem{Bergin et al. (2007)}  Bergin, E.~A., Aikawa, Y., Blake, G.~A., \& van Dishoeck, E.~F.,  {\it "The Chemical Evolution of Protoplanetary Disks"},  Protostars and Planets V,   751,  2007

\bibitem{Berthoud et al.(2007)}  Berthoud, M.~G., Keller, L.~D., Herter, 
T.~L., Richter, M.~J., 
\& Whelan, D.~G.,  {\it "Near-IR CO Overtone Emission in 51 Ophiuchi"},  Astrophysical Journal,  660,  461,  2007 

\bibitem{Bitner et al.(2007)}  Bitner, M.~A., Richter, M.~J., Lacy, J.~H., 
Greathouse, T.~K., Jaffe, D.~T., \& Blake, G.~A.,  {\it "TEXES Observations of Pure Rotational H$_2$ Emission from AB Aurigae"},  Astrophysical Journal,  661,  L69,  2007

\bibitem{Bitner et al.(2008)}  Bitner, M.~A., et al.,  {\it "The TEXES 
Survey For H2 Emission From Protoplanetary Disks"},  ArXiv e-prints,   
arXiv:0808.1099,  2008 

\bibitem{Blake(2003)}  Blake, G.~A.,  {\it "Chemistry in Circumstellar 
Disks as Probed by High Resolution Millimeter-wave to Infrared 
Spectroscopy"},  SFChem 2002: Chemistry as a Diagnostic of Star Formation,   
178,  2003 

\bibitem{Blake et al.(2004)}  Blake, G.~A., \& Boogert, A.~C.~A.,  {\it "High-Resolution 4.7 Micron Keck/NIRSPEC Spectroscopy of the CO Emission from the Disks Surrounding Herbig Ae Stars"},  Astrophysical Journal,  606,  L73,  2004

\bibitem{Brittain et al.(2002)}  Brittain, S.~D., \& Rettig, T.~W.,  {\it "CO and H$_3^+$ in the protoplanetary disk around the star HD141569"},  Nature,  418,  57,  2002

\bibitem{Brittain et al.(2003)}  Brittain, S.~D., Rettig, T.~W., Simon, T., 
Kulesa, C., DiSanti, M.~A., \& Dello Russo, N.,  {\it "CO Emission from Disks around AB Aurigae and HD 141569: Implications for Disk Structure and Planet Formation Timescales"},  Astrophysical Journal,  588,  535,  2003 

\bibitem{Brittain et al. (2005)}  Brittain, S.~D., Rettig, T.~W., Simon, T., \& Kulesa, C.,  {\it "CO Line Emission and Absorption from the HL Tauri Disk-Where Is All the Dust?"},  Astrophysical Journal,  626,  283,  2005 

\bibitem{Brittain et al. (2007a)}  Brittain, S.~D., Simon, T., Najita, J.~R., \& Rettig, T.~W.,  {\it "Warm Gas in the Inner Disks around Young Intermediate-Mass Stars"},  Astrophysical Journal,  659,  685,  2007a

\bibitem{Brittain et al.(2007b)}  Brittain, S., Rettig, T.~W., Simon, T., 
Balsara, D.~S., Tilley, D., Gibb, E., \& Hinkle, K.~H.,  {\it "Post-Outburst Observations of V1647 Orionis: Detection of a Brief Warm Molecular Outflow"},  Astrophysical Journal,  670,  L29,  2007b 

\bibitem{Brown et al. (2005)}  Brown, J.~M., Boogert, A.~C.~A., Salyk, C., \& Blake, G.~A.,  {\it "High Resolution 4.7 
$\mu$m Keck/NIRSPEC Spectra of Protostars"},  Protostars and Planets V,   8513,  2005

\bibitem{Bouret et al.(2003)}  Bouret, J.-C., Martin, C., Deleuil, M., 
Simon, T., 
\& Catala, C.,  {\it "Far UV spectroscopy of the circumstellar environment of the Herbig Be stars HD 259431 and HD 250550"},  Astronomy and Astrophysics,  410,  175,  2003 

\bibitem{Calvet et al.(2005)}  Calvet, N., et al.,  {\it "Disks in 
Transition in the Taurus Population: Spitzer IRS Spectra of GM Aurigae and 
DM Tauri"},  Astrophysical Journal,  630,  L185,  2005 

\bibitem{Carmona et al.(2005)}  Carmona, A., van den Ancker, M.~E., Thi, 
W.-F., Goto, M., 
\& Henning, T.,  {\it "Upper limits on CO 4.7 $\mu$m emission from disks around five Herbig Ae/Be stars"},  Astronomy and Astrophysics,  436,  977,  2005

\bibitem{Carmona et al.(2007)}  Carmona, A., van den Ancker, M.~E., 
Henning, T., Goto, M., Fedele, D., 
\& Stecklum, B.,  {\it "A search for near-infrared molecular hydrogen emission in the CTTS LkH$\alpha$; 264 and the debris disk 49 Ceti"},  Astronomy and Astrophysics,  476,  853,  2007

\bibitem{Carmona et al.(2008a)}  Carmona, A., van den Ancker, M.~E., 
Henning, T., Goto, M., Fedele, D., 
\& Stecklum, B.,  {\it "Erratum: A search for near-infrared molecular hydrogen emission in the CTTS LkH$\alpha$ 264 and the debris disk 49 Ceti"},  Astronomy and Astrophysics,  478,  795,  2008a

\bibitem{Carmona et al.(2008b)}  Carmona, A., et al.,  {\it "A search for 
mid-infrared molecular hydrogen emission from protoplanetary disks"},  
Astronomy and Astrophysics,  477,  839,  2008b 

\bibitem{Carr(1989)}  Carr, J.~S.,  {\it "Near-infrared CO emission in 
young stellar objects"},  Astrophysical Journal,  345,  522,  1989 

\bibitem{Carr and Tokunaga(1992)}  Carr, J.~S., \& Tokunaga, A.~T.,  {\it "Measurement of CO overtone line profiles in SVS 13"},  Astrophysical Journal,  393,  L67,  1992 

\bibitem{Carr et al.(1993)}  Carr, J.~S., Tokunaga, A.~T., Najita, J., Shu, 
F.~H., \& Glassgold, A.~E.,  {\it "The inner-disk and stellar properties of the young stellar object WL 16"},  Astrophysical Journal,  411,  L37,  1993 

\bibitem{Carr et al. (2001)}  Carr, J.~S., Mathieu, R.~D., \& Najita, J.~R.,  {\it "Evidence for Residual Material in Accretion Disk Gaps: CO Fundamental Emission from the T Tauri Spectroscopic Binary DQ Tauri"},  Astrophysical Journal,  551,  454,  2001

\bibitem{Carr et al.(2004)}  Carr, J.~S., Tokunaga, A.~T., \& Najita, J.,  {\it "Hot H$_2$O Emission and Evidence for Turbulence in the Disk of a Young Star"},  Astrophysical Journal,  603,  213,  2004 

\bibitem{Carr(2005)}  Carr, J.,  {\it "High-Resolution Infrared 
Spectroscopy of Protoplanetary Disks"},  in High Resolution Infrared 
Spectroscopy in Astronomy, Springer Verlag, 203,  2005 

\bibitem{Carr and Najita (2008)}  Carr, J.~S., \& Najita, J.~R.,  {\it "Organic Molecules and Water in the Planet Formation Region of Young Circumstellar Disks"},  Science,  319,  1504,  2008

\bibitem{Chandler et al.(1993)}  Chandler, C.~J., Carlstrom, J.~E., 
Scoville, N.~Z., Dent, W.~R.~F., \& Geballe, T.~R.,  {\it "Infrared CO emission from young stars - High-resolution spectroscopy"},  Astrophysical Journal,  412,  L71,  1993  

\bibitem{Chiang and Goldreich(1997)}  Chiang, E.~I., \& Goldreich, P.,  {\it "Spectral Energy Distributions of T Tauri Stars with Passive Circumstellar Disks"},  Astrophysical Journal, 490, 368, 1997

\bibitem{Cieza et al. (2008)}  Cieza, L.~A., Swift, J.~J., Mathews, G.~S., \& Williams, J.~P.,  {\it "The Masses of Transition Circumstellar Disks: Observational Support for Photoevaporation Models"},  Astrophysical Journal,  686,  L115,  2008 

\bibitem{Dent et al. (2005)}  Dent, W.~R.~F., Greaves, J.~S., \& Coulson, I.~M.,  {\it "CO emission from discs around isolated HAeBe and Vega-excess stars"},  Monthly Notices of the Royal Astronomical Society,  359,  663,  2005

\bibitem{Doppmann et al. (2008)}  Doppmann, G.~W., Najita, J.~R., \& Carr, J.~S.,  {\it "Stellar and Circumstellar Properties of the Pre-Main-Sequence Binary GV Tau from Infrared Spectroscopy"},  Astrophysical Journal,  685,  298,  2008 

\bibitem{Dutrey et al.(1996)}  Dutrey, A., Guilloteau, S., Duvert, G., 
Prato, L., Simon, M., Schuster, K., 
\& Menard, F.,  {\it "Dust and gas distribution around T Tauri stars in Taurus-Auriga. I. Interferometric 2.7mm continuum and $^{13}$CO J=1-0 observations"},  Astronomy and Astrophysics,  309,  493,  1996 

\bibitem{Dutrey et al.(1998)}  Dutrey, A., Guilloteau, S., Prato, L., Simon, M., Duvert, G., Schuster, K., 
\& Menard, F.,  {\it "CO study of the GM Aurigae Keplerian disk"},  Astronomy and Astrophysics,  338,  L63,  1998 

\bibitem{Dutrey, Guilloteau, and Ho(2007)}  Dutrey, A., Guilloteau, S., \& Ho, P.,  {\it "Interferometric Spectroimaging of Molecular Gas in Protoplanetary Disks"},  Protostars and Planets V,   495,  2007 

\bibitem{Errico et al. (2001)}  Errico, L., Lamzin, S.~A., \& Vittone, A.~A.,  {\it "UV spectra of T Tauri stars from the HST and IUE satellites: BP Tau"},  Astronomy and Astrophysics,  377,  557,  2001 

\bibitem{Espaillat et al.(2007)}  Espaillat, C., et al.,  {\it "Probing the 
Dust and Gas in the Transitional Disk of CS Cha with Spitzer"},  
Astrophysical Journal,  664,  L111,  2007 

\bibitem{Fedele et al.(2008)}  Fedele, D., et al.,  {\it "The structure of 
the protoplanetary disk surrounding three young intermediate mass stars. 
II. Spatially resolved dust and gas distribution"},  ArXiv e-prints,   
arXiv:0809.3947,  2008 

\bibitem{France et al.(2007)}  France, K., Roberge, A., Lupu, R.~E., 
Redfield, S., \& Feldman, P.~D.,  {\it "A Low-Mass H$_2$ Component to the AU Microscopii Circumstellar Disk"},  Astrophysical Journal,  668,  1174,  2007

\bibitem{Geballe and Persson(1987)}  Geballe, T.~R., \& Persson, S.~E.,  {\it "Emission from CO band heads in young stellar objects"},  Astrophysical Journal,  312,  297,  1987 

\bibitem{Gibb et al. (2007)}  Gibb, E.~L., Van Brunt, K.~A., Brittain, S.~D., \& Rettig, T.~W.,  {\it "Warm HCN, C$_2$H$_2$, and CO in the Disk of GV Tau"},  Astrophysical Journal,  660,  1572,  2007 

\bibitem{Gibb et al. (2008)}  Gibb, E.~L., Van Brunt, K.~A., Brittain, S.~D., \& Rettig, T.~W.,  {\it "Erratum: ``Warm HCN, C$_2$H$_2$ and CO in the Disk of GV Tau''},  Astrophysical Journal,  686,  748,  2008 

\bibitem{Gizis et al. (2005)}  Gizis, J.~E., Shipman, H.~L., \& Harvin, J.~A.,  {\it "First Ultraviolet Spectrum of a Brown Dwarf: Evidence for $H_2$ Fluorescence and Accretion"},  Astrophysical Journal,  630,  L89,  2005

\bibitem{Glassgold et al. (2007)}  Glassgold, A.~E., Najita, J.~R., \& Igea, J.,  {\it "Neon Fine-Structure Line Emission by X-Ray Irradiated Protoplanetary Disks"},  Astrophysical Journal,  656,  515,  2007 

\bibitem{Gorti and Hollenbach(2008)}  Gorti, U., \& Hollenbach, D.,  {\it "Line Emission from Gas in Optically Thick Dust Disks around Young Stars"},  Astrophysical Journal,  683,  287,  2008 

\bibitem{Grady et al.(2005)}  Grady, C.~A., et al.,  {\it "Coronagraphic 
Imaging of Pre-Main-Sequence Stars with the Hubble Space Telescope Space 
Telescope Imaging Spectrograph. I. The Herbig Ae Stars"},  Astrophysical 
Journal,  630,  958,  2005

\bibitem{Gustafsson et al. (2008)}  Gustafsson, M., Labadie, L., Herbst, T.~M., \& Kasper, M.,  {\it "Spatially resolved H$_2$ emission from the disk around T Tau N"},  Astronomy and Astrophysics,  488,  235,  2008  

\bibitem{Guilloteau and Dutrey(1994)}  Guilloteau, S., \& Dutrey, A.,  {\it "A Keplerian disk around DM Tau?"},  Astronomy and Astrophysics,  291,  L23,  1994

\bibitem{Guilloteau and Dutrey(1998)}  Guilloteau, S., \& Dutrey, A.,  {\it "Physical parameters of the Keplerian protoplanetary disk of DM Tauri"},  Astronomy and Astrophysics,  339,  467,  1998 

\bibitem{Goto et al.(2006)}  Goto, M., Usuda, T., Dullemond, C.~P., 
Henning, T., Linz, H., Stecklum, B., \& Suto, H.,  {\it "Inner Rim of a Molecular Disk Spatially Resolved in Infrared CO Emission Lines"},  Astrophysical Journal,  652,  758,  2006 

\bibitem{Haisch et al. (2001)} Haisch, K.~E., Jr., Lada, E.~A., \& Lada, C.~J.,  {\it "Disk Frequencies and Lifetimes in Young Clusters"},  Astrophysical Journal,  553,  L153,  2001

\bibitem{Hartmann et al. (1994)}  Hartmann, L., Hewett, R., \& Calvet, N.,  {\it "Magnetospheric accretion models for T Tauri stars. 1: Balmer line profiles without rotation"},  Astrophysical Journal,  426,  669,  1994

\bibitem{Herczeg et al.(2002)}  Herczeg, G.~J., Linsky, J.~L., Valenti, 
J.~A., Johns-Krull, C.~M., 
\& Wood, B.~E.,  {\it "The Far-Ultraviolet Spectrum of TW Hydrae. I. Observations of $H_2$ Fluorescence"},  Astrophysical Journal,  572,  310,  2002 

\bibitem{Herczeg et al.(2004)}  Herczeg, G.~J., Wood, B.~E., Linsky, J.~L., 
Valenti, J.~A., 
\& Johns-Krull, C.~M.,  {\it "The Far-Ultraviolet Spectra of TW Hydrae. II. Models of H$_2$ Fluorescence in a Disk"},  Astrophysical Journal,  607,  369,  2004 

\bibitem{Herczeg et al.(2005)}  Herczeg, G.~J., et al.,  {\it "The Loopy 
Ultraviolet Line Profiles of RU Lupi: Accretion, Outflows, and 
Fluorescence"},  Astronomical Journal,  129,  2777,  2005 

\bibitem{Herczeg et al.(2006)}  Herczeg, G.~J., Linsky, J.~L., Walter, F.~M., Gahm, G.~F., 
\& Johns-Krull, C.~M.,  {\it "The Origins of Fluorescent H$_2$ Emission From T Tauri Stars"},  Astrophysical Journal Supplement Series,  165,  256,  2006

\bibitem{Herczeg et al.(2007)}  Herczeg, G.~J., Najita, J.~R., Hillenbrand, L.~A., \& Pascucci, I.,  {\it "High-Resolution Spectroscopy of [Ne II] Emission from TW Hydrae"},  Astrophysical Journal,  670,  509,  2007

\bibitem{Henning(2006)}  Henning, Th.,  {\it "Dust in Protoplanetary 
Disks"},  Meteoritics \& Planetary Science, Vol. 41, Supplement, Proceedings 
of 69th Annual Meeting of the Meteoritical Society, held August 6-11, 2006 
in Zurich, Switzerland., p.5392,  41,  5392,  2006 

\bibitem{Hollenbach and McKee(1989)}  Hollenbach, D., \& McKee, C.~F.,  {\it "Molecule formation and infrared emission in fast interstellar shocks. III - Results for J shocks in molecular clouds"},  Astrophysical Journal,  342,  306,  1989

\bibitem{Hollenbach et al.(2005)}  Hollenbach, D., et al.,  {\it "Formation 
and Evolution of Planetary Systems: Upper Limits to the Gas Mass in HD 
105"},  Astrophysical Journal,  631,  1180,  2005 

\bibitem{Itoh et al.(2003)}  Itoh, Y., Sugitani, K., Ogura, K., \& Tamura, M.,  {\it "Detection of Molecular Hydrogen Emission Associated with LkH$\alpha$ 264"},  Publications of the Astronomical Society of Japan,  55,  L77,  2003

\bibitem{Koerner et al. (1993)}  Koerner, D.~W., Sargent, A.~I., \& Beckwith, S.~V.~W.,  {\it "A rotating gaseous disk around the T Tauri star GM Aurigae"},  Icarus,  106,  2,  1993

\bibitem{Lahuis et al.(2006)}  Lahuis, F., et al.,  {\it "Hot Organic 
Molecules toward a Young Low-Mass Star: A Look at Inner Disk Chemistry"},  
Astrophysical Journal,  636,  L145,  2006

\bibitem{Lahuis et al.(2007)}  Lahuis, F., van Dishoeck, E.~F., Blake, 
G.~A., Evans, N.~J., II, Kessler-Silacci, J.~E., 
\& Pontoppidan, K.~M.,  {\it "c2d Spitzer IRS Spectra of Disks around T Tauri Stars. III. [Ne II], [Fe I], and H$_2$ Gas-Phase Lines"},  Astrophysical Journal,  665,  492,  2007

\bibitem{Lecavelier des Etangs et al.(2001)}  Lecavelier des Etangs, A., et 
al.,  {\it "Deficiency of molecular hydrogen in the disk of $\beta$ Pictoris"},  
Nature,  412,  706,  2001 

\bibitem{Lecavelier des Etangs et al.(2003)}  Lecavelier des Etangs, A., et 
al.,  {\it "FUSE observations of H$_2$ around the Herbig AeBe stars  HD 
100546 and HD 163296"},  Astronomy and Astrophysics,  407,  935,  2003 

\bibitem{Mandell et al.(2008)}  Mandell, A.~M., Mumma, M.~J., Blake, G.~A., 
Bonev, B.~P., Villanueva, G.~L., 
\& Salyk, C.,  {\it "Discovery of OH in Circumstellar Disks around Young Intermediate-Mass Stars"},  Astrophysical Journal,  681,  L25,  2008 

\bibitem{Mannings and Sargent(1997)}  Mannings, V., \& Sargent, A.~I.,  {\it "A High-Resolution Study of Gas and Dust around Young Intermediate-Mass Stars: Evidence for Circumstellar Disks in Herbig AE Systems"},  Astrophysical Journal,  490,  792,  1997

\bibitem{Martin et al.(2004)}  Martin, C., Bouret, J.-C., Deleuil, M., 
Simon, T., 
\& Catala, C.,  {\it "Far Ultraviolet Spectroscopy of HD 76534"},  Astronomy and Astrophysics,  416,  L5,  2004 

\bibitem{Martin-Zaidi et al.(2005)}  Martin-Za{\"i}di, C., Deleuil, M., 
Simon, T., Bouret, J.-C., Roberge, A., Feldman, P.~D., Lecavelier Des 
Etangs, A., \& Vidal-Madjar, A.,  {\it "FUSE observations of molecular hydrogen on the line of sight towards HD 141569A"},  Astronomy and Astrophysics,  440,  921,  2005

\bibitem{Martin-Zaidi et al. (2007)}  Martin-Za{\"i}di, C., Lagage, P.-O., Pantin, E., \& Habart, E.,  {\it "Detection of Warm Molecular Hydrogen in the Circumstellar Disk around the Herbig Ae Star HD 97048"},  Astrophysical Journal,  666,  L117,  2007 

\bibitem{Martin-Zaidi et al.(2008)}  Martin-Za{\"i}di, C., et al.,  
{\it "Molecular hydrogen in the circumstellar environments of Herbig Ae/Be 
stars probed by FUSE"},  Astronomy and Astrophysics,  484,  225,  2008a 

\bibitem{Martin-Zaidi et al.(2008)}  Martin-Za{\"i}di, C., van 
Dishoeck, E.~F., Augereau, J.-C., Lagage, P.-O., 
\& Pantin, E.,  {\it "Searching for molecular hydrogen mid-infrared emission in the circumstellar environments of Herbig Be stars"},  Astronomy and Astrophysics,  489,  601,  2008b

\bibitem{Meeus et al.(2001)}  Meeus, G., Waters, L.~B.~F.~M., Bouwman, J., 
van den Ancker, M.~E., Waelkens, C., \& Malfait, K.,  {\it "ISO spectroscopy of circumstellar dust in 14 Herbig Ae/Be systems: Towards an understanding of dust processing"},  Astronomy and Astrophysics,  365,  476,  2001 

\bibitem{Meijerink et al.(2008)}  Meijerink, R., Glassgold, A.~E., \& Najita, J.~R.,  {\it "Atomic Diagnostics of X-Ray-Irradiated Protoplanetary Disks"},  Astrophysical Journal,  676,  518,  2008 

\bibitem{Muzerolle et al. (1998a)}  Muzerolle, J., Hartmann, L., \& Calvet, N.,  {\it "A BR gamma Probe of Disk Accretion in T Tauri Stars and Embedded Young Stellar Objects"},  Astronomical Journal,  116,  2965,  1998a 

\bibitem{Muzerolle et al. (1998b)}  Muzerolle, J., Calvet, N., \& Hartmann, L.,  {\it "Magnetospheric Accretion Models for the Hydrogen Emission Lines of T Tauri Stars"},  Astrophysical Journal,  492,  743,  1998b 

\bibitem{Muzerolle et al. (2003)}  Muzerolle, J., Calvet, N., Hartmann, L., \& D'Alessio, P.,  {\it "Unveiling the Inner Disk Structure of T Tauri Stars"},  Astrophysical Journal,  597,  L149,  2003 

\bibitem{Najita et al. (1996a)}  Najita, J., Carr, J.~S., \& Tokunaga, A.~T.,  {\it "High-Resolution Spectroscopy of BR gamma Emission in Young Stellar Objects"},  Astrophysical Journal,  456,  292,  1996a 

\bibitem{Najita et al.(1996b)}  Najita, J., Carr, J.~S., Glassgold, A.~E., 
Shu, F.~H., \& Tokunaga, A.~T.,  {\it "Kinematic Diagnostics of Disks around Young Stars: CO Overtone Emission from WL 16 and 1548C27"},  Astrophysical Journal,  462,  919,  1996b 

\bibitem{Najita et al. (2000)}  Najita, J.~R., Edwards, S., Basri, G., \& Carr, J.,  {\it "Spectroscopy of Inner Protoplanetary Disks and the Star-Disk Interface"},  Protostars and Planets IV,   457,  2000 

\bibitem{Najita et al. (2003)}  Najita, J., Carr, J.~S., \& Mathieu, R.~D.,  {\it "Gas in the Terrestrial Planet Region of Disks: CO Fundamental Emission from T Tauri Stars"},  Astrophysical Journal,  589,  931,  2003

\bibitem{Najita et al. (2007a)}  Najita, J.~R., Carr, J.~S., Glassgold, A.~E., \& Valenti, J.~A.,  {\it "Gaseous Inner Disks"},  Protostars and Planets V,   507,  2007a 

\bibitem{Najita et al. (2007b)}  Najita, J.~R., Strom, S.~E., \& Muzerolle, J.,  {\it "Demographics of transition objects"},  Monthly Notices of the Royal Astronomical Society,  378,  369,  2007b 

\bibitem{Najita et al. (2008)}  Najita, J.~R., Crockett, N., \& Carr, J.~S.,  {\it "CO Fundamental Emission from V836 Tau"},  ArXiv e-prints,   arXiv:0809.3949,  2008 

\bibitem{Natta et al. (2006)}  Natta, A., Testi, L., \& Randich, S.,  {\it "Accretion in the $\rho$-Ophiuchi pre-main sequence stars"},  Astronomy and Astrophysics,  452,  245,  2006

\bibitem{Natta et al.(2007)}  Natta, A., Testi, L., Calvet, N., Henning, 
T., Waters, R., 
\& Wilner, D.,  {\it "Dust in Protoplanetary Disks: Properties and Evolution"},  Protostars and Planets V,   767,  2007 

\bibitem{Nomura et al. (2005)}  Nomura, H., \& Millar, T.~J.,  {\it "Molecular hydrogen emission from protoplanetary disks"},  Astronomy and Astrophysics,  438,  923,  2005
 
\bibitem{Nomura et al.(2007)}  Nomura, H., Aikawa, Y., Tsujimoto, M., 
Nakagawa, Y., \& Millar, T.~J.,  {\it "Molecular Hydrogen Emission from Protoplanetary Disks. II. Effects of X-Ray Irradiation and Dust Evolution"},  Astrophysical Journal,  661,  334,  2007 

\bibitem{Pascucci et al.(2006)}  Pascucci, I., et al.,  {\it "Formation and 
Evolution of Planetary Systems: Upper Limits to the Gas Mass in Disks 
around Sun-like Stars"},  Astrophysical Journal,  651,  1177,  2006 

\bibitem{Pascucci et al.(2007)}  Pascucci, I., et al.,  {\it "Detection of 
[Ne II] Emission from Young Circumstellar Disks"},  Astrophysical Journal,  
663,  383,  2007 

\bibitem{Pietu et al. (2007)}  Pi\'etu, V., Dutrey, A., \& Guilloteau, S.,  {\it "Probing the structure of protoplanetary disks: a comparative study of DM Tau, LkCa 15, and MWC 480"},  Astronomy and Astrophysics,  467,  163,  2007

\bibitem{Pontoppidan et al.(2008)}  Pontoppidan, K.~M., Blake, G.~A., van 
Dishoeck, E.~F., Smette, A., Ireland, M.~J., 
\& Brown, J.,  {\it "Spectroastrometric Imaging of Molecular Gas within Protoplanetary Disk Gaps"},  Astrophysical Journal,  684,  1323,  2008 

\bibitem{Ramsay Howat et al. (2007)}  Ramsay Howat, S.~K., \& Greaves, J.~S.,  {\it "Molecular hydrogen emission from discs in the $\eta$; Chamaeleontis cluster"},  Monthly Notices of the Royal Astronomical Society,  379,  1658,  200

\bibitem{Rettig et al.(2004)}  Rettig, T.~W., Haywood, J., Simon, T., Brittain, S.~D., 
\& Gibb, E.,  {\it "Discovery of CO Gas in the Inner Disk of TW Hydrae"},  Astrophysical Journal,  616,  L163,  2004

\bibitem{Rettig et al.(2005)}  Rettig, T.~W., Brittain, S.~D., Gibb, E.~L., 
Simon, T., 
\& Kulesa, C.,  {\it "CO Emission and Absorption toward V1647 Orionis (McNeil's Nebula)"},  Astrophysical Journal,  626,  245,  2005 

\bibitem{Richter et al. (2002)}  Richter, M.~J., Jaffe, D.~T., Blake, G.~A., \& Lacy, J.~H.,  {\it "Looking for Pure Rotational H$_2$ Emission from Protoplanetary Disks"},  Astrophysical Journal,  572,  L161,  2002 

\bibitem{Roberge et al.(2001)}  Roberge, A., et al.,  {\it "FUSE and Hubble 
Space Telescope/STIS Observations of Hot and Cold Gas in the AB Aurigae 
System"},  Astrophysical Journal,  551,  L97,  2001 

\bibitem{Roberge et al.(2005)}  Roberge, A., Weinberger, A.~J., Redfield, S., \& Feldman, P.~D.,  {\it "Rapid Dissipation of Primordial Gas from the AU Microscopii Debris Disk"},  Astrophysical Journal,  626,  L105,  2005  


\bibitem{Sako et al.(2005)}  Sako, S., Yamashita, T., Kataza, H., Miyata, 
T., Okamoto, Y.~K., Honda, M., Fujiyoshi, T., 
\& Onaka, T.,  {\it "Search for 17 $\mu$m H$_2$ Pure Rotational Emission from Circumstellar Disks"},  Astrophysical Journal,  620,  347,  2005 

\bibitem{Salyk et al. (2007)}  Salyk, C., Blake, G.~A., Boogert, A.~C.~A., \& Brown, J.~M.,  {\it "Molecular Gas in the Inner 1 AU of the TW Hya and GM Aur Transitional Disks"},  Astrophysical Journal,  655,  L105,  2007 

\bibitem{Salyk et al.(2008)}  Salyk, C., Pontoppidan, K.~M., Blake, G.~A., 
Lahuis, F., van Dishoeck, E.~F., 
\& Evans, N.~J., II,  {\it "H$_2$O and OH Gas in the Terrestrial Planet-forming Zones of Protoplanetary Disks"},  Astrophysical Journal,  676,  L49,  2008

\bibitem{Sheret et al. (2003)}  Sheret, I., Ramsay Howat, S.~K., \& Dent, W.~R.~F.,  {\it "A search for H$_2$ around pre-main-sequence stars"},  Monthly Notices of the Royal Astronomical Society,  343,  L65,  2003 

\bibitem{Sicilia-Aguilar et al.(2006)}  Sicilia-Aguilar, A., Hartmann, 
L.~W., F{\"u}r{\'e}sz, G., Henning, T., Dullemond, C., 
\& Brandner, W.,  {\it "High-Resolution Spectroscopy in Tr 37: Gas Accretion Evolution in Evolved Dusty Disks"},  Astronomical Journal,  132,  2135,  2006

\bibitem{Scoville et al.(1983)}  Scoville, N., Kleinmann, S.~G., Hall, D.~N.~B., \& Ridgway, S.~T.,  {\it "The circumstellar and nebular environment of the Becklin-Neugebauer object - 2-5 micron wavelength spectroscopy"},  Astrophysical Journal,  275,  201,  1983 

\bibitem{Shukla et al. (2003)}  Shukla, S.~J., Bary, J.~S., Weintraub, D.~A., \& Kastner, J.~H.,  {\it "Further detections of rovibrational H$_2$ emission from the disks of T Tauri stars"},  Bulletin of the American Astronomical Society,  35,  1209,  2003 

\bibitem{Skrutskie et al.(1993)}  Skrutskie, M.~F., et al.,  {\it 
"Detection of circumstellar gas associated with GG Tauri"},  Astrophysical 
Journal,  409,  422,  1993

\bibitem{Tatulli et al.(2008)}  Tatulli, E., et al.,  {\it "Spatially 
resolving the hot CO around the young Be star 51 Ophiuchi"},  Astronomy and 
Astrophysics,  489,  1151,  2008

\bibitem{Thi et al.(2001)}  Thi, W.~F., et al.,  {\it "H$_2$ and CO Emission 
from Disks around T Tauri and Herbig Ae Pre-Main-Sequence Stars and from 
Debris Disks around Young Stars: Warm and Cold Circumstellar Gas"},  
Astrophysical Journal,  561,  1074,  2001

\bibitem{Thi et al. (2004)}  Thi, W.-F., van Zadelhoff, G.-J., \& van Dishoeck, E.~F.,  {\it "Organic molecules in protoplanetary disks around T Tauri and Herbig Ae stars"},  Astronomy and Astrophysics,  425,  955,  2004

\bibitem{Thi et al. (2005a)}  Thi, W.-F., van Dalen, B., Bik, A., \& Waters, L.~B.~F.~M.,  {\it "Evidence for a hot dust-free inner disk around 51 Oph"},  Astronomy and Astrophysics,  430,  L61,  2005a 

\bibitem{Thi and Bik (2005b)}  Thi, W.-F., \& Bik, A.,  {\it "Detection of steam in the circumstellar disk around a massive Young Stellar Object"},  Astronomy and Astrophysics,  438,  557,  2005b

\bibitem{Valenti et al.(2000)}  Valenti, J.~A., Johns-Krull, C.~M., \& Linsky, J.~L.,  {\it "An IUE Atlas of Pre-Main-Sequence Stars. I. Co-added Final Archive Spectra from the SWP Camera"},  Astrophysical Journal Supplement Series,  129,  399,  2000 

\bibitem{van den Ancker et al.(1999)}  van den Ancker, M.~E., Wesselius, 
P.~R., Tielens, A.~G.~G.~M., van Dishoeck, E.~F., 
\& Spinoglio, L.,  {\it "ISO spectroscopy of shocked gas in the vicinity of T Tauri"},  Astronomy and Astrophysics,  348,  877,  1999 

\bibitem{van der Plas et al.(2008a)}  van der Plas, G., van den Ancker, 
M.~E., Fedele, D., Acke, B., Dominik, C., Waters, L.~B.~F.~M., 
\& Bouwman, J.,  {\it "The structure of protoplanetary disks surrounding three young intermediate mass stars. I. Resolving the disk rotation in the [OI] 6300 \AA~ line"},  Astronomy and Astrophysics,  485,  487,  2008a

\bibitem{van der Plas et al.(2008b)}  van der Plas, G., van den Ancker, 
M.~E., Acke, B., Carmona, A., Dominik, C., Fedele, D., 
\& Waters, L.~B.~F.~M.,  {\it "Spatially resolved 4.7 $\mu$m CO fundamental emission in two protoplanetary disks"},  ArXiv e-prints,   arXiv:0810.3417,  2008b

\bibitem{van Langevelde et al. (1994)}  van Langevelde, H.~J., van Dishoeck, E.~F., van der Werf, P.~P., \& Blake, G.~A.,  {\it "The spatial distribution of excited H2 in T Tau: A molecular outflow in a young binary system"},  Astronomy and Astrophysics,  287,  L25,  1994 

\bibitem{Walter et al.(2003)}  Walter, F.~M., et al.,  {\it "Mapping the 
Circumstellar Environment of T Tauri with Fluorescent H$_2$ Emission"},  
Astronomical Journal,  126,  3076,  2003 

\bibitem{Weintraub et al.(1989)}  Weintraub, D.~A., Zuckerman, B., \& Masson, C.~R.,  {\it "Measurements of Keplerian rotation of the gas in the circumbinary disk around T Tauri"},  Astrophysical Journal,  344,  915,  1989 

\bibitem{Weintraub, Kastner and Bary (2000)}  Weintraub, D.~A., Kastner, J.~H., \& Bary, J.~S.,  {\it "Detection of Quiescent Molecular Hydrogen Gas in the Circumstellar Disk of a Classical T Tauri Star"},  Astrophysical Journal,  541,  767,  2000 

\bibitem{Weintraub et al.(2005)} Weintraub, D.~A., 
Bary, J.~S., Kastner, J.~H., Shukla, S.~J., 
\& Chynoweth, K.\ 2005, Bulletin of the American Astronomical Society, 37, 1165 

\bibitem{Wilkinson et al. (2002)}  Wilkinson, E., Harper, G.~M., Brown, A., \& Herczeg, G.~J.,  {\it "The Far-Ultraviolet Spectrum of T Tauri between 912 and 1185 \AA"},  Astronomical Journal,  124,  1077,  2002 



\end{thebibliography}
\end{document}